\documentclass[manuscript]{vgtc}                          % final (conference style)

\graphicspath{{figures/}{pictures/}{images/}{./}} % where to search for the images

\usepackage{multirow}
\usepackage[normalem]{ulem}
\usepackage{tabularx}
\usepackage[table,xcdraw]{xcolor}

\usepackage{bbm}
\usepackage{soul}
\usepackage{times}                     % we use Times as the main font
\usepackage{tabu}                      % only used for the table example
\usepackage{booktabs}                  % only used for the table example
\usepackage{lipsum}                    % used to generate placeholder text
\usepackage{mwe}                       % used to generate placeholder figures

\usepackage{mathptmx}                  % use matching math font

\usepackage{balance}

\usepackage{xcolor}
 
\usepackage[final]{changes} 
\makeatletter
\setdeletedmarkup{\@gobble{#1}}
\makeatother

\onlineid{0}

\vgtccategory{Research}

\vgtcinsertpkg

\title{Analyzing Multimodal Interaction Strategies for LLM-Assisted Manipulation of 3D Scenes}

\author{Junlong Chen\thanks{e-mail: jc2375@cam.ac.uk}\\ %
        \scriptsize University of Cambridge %
\and Jens Grubert \thanks{e-mail: jens.grubert@hs-coburg.de}\\ %
     \scriptsize Coburg University of Applied Sciences %
\and Per Ola Kristensson \thanks{e-mail: pok21@cam.ac.uk}\\ %
     \parbox{1.4in}{\scriptsize \centering University of Cambridge}}

\teaser{
  \centering
  \includegraphics[width=0.329\textwidth]{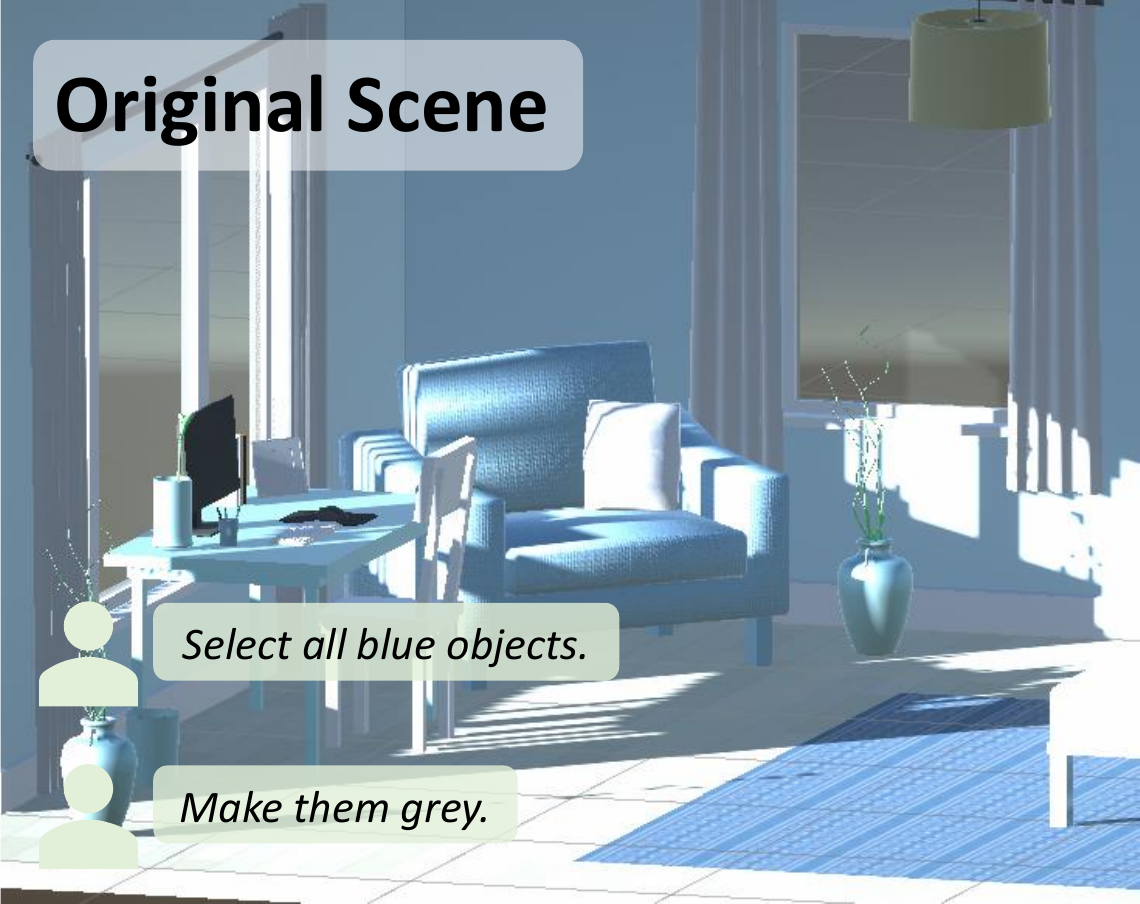}
  \includegraphics[width=0.329\textwidth]{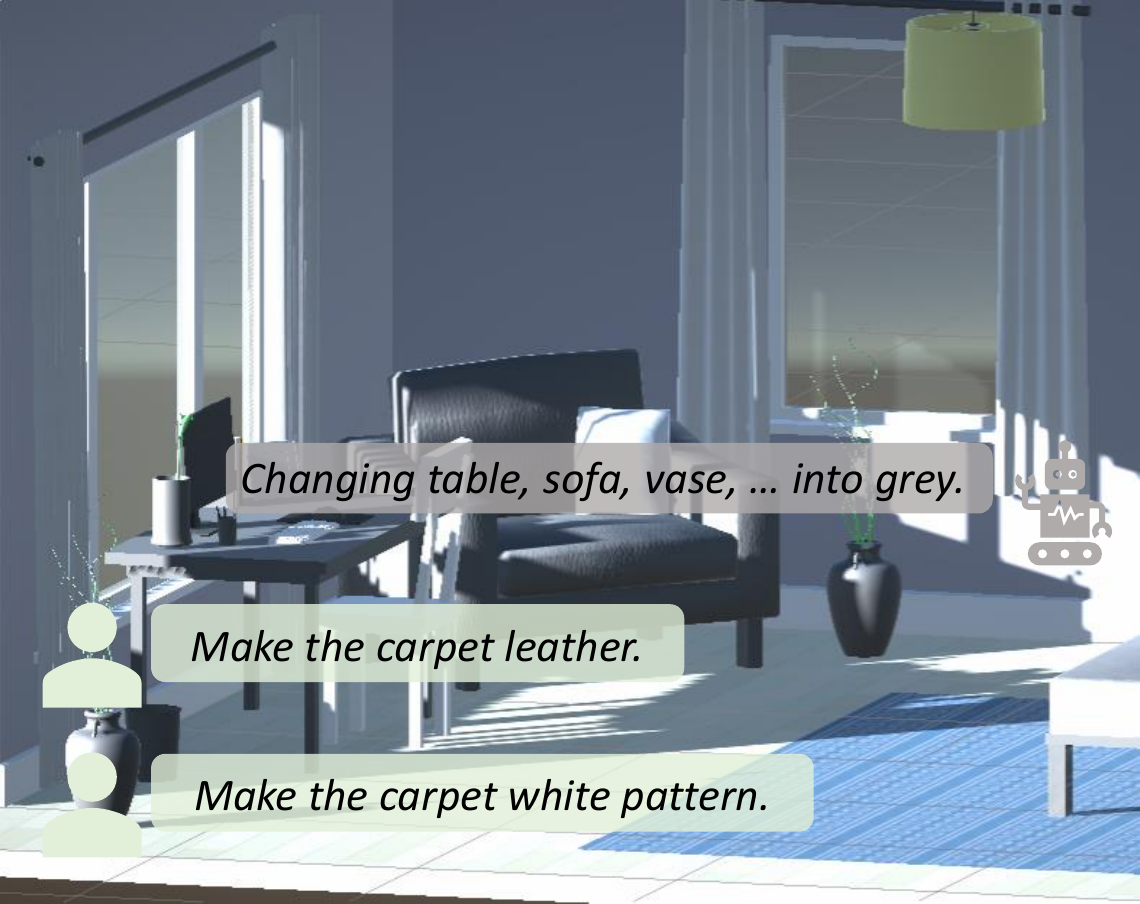}
  \includegraphics[width=0.329\textwidth]{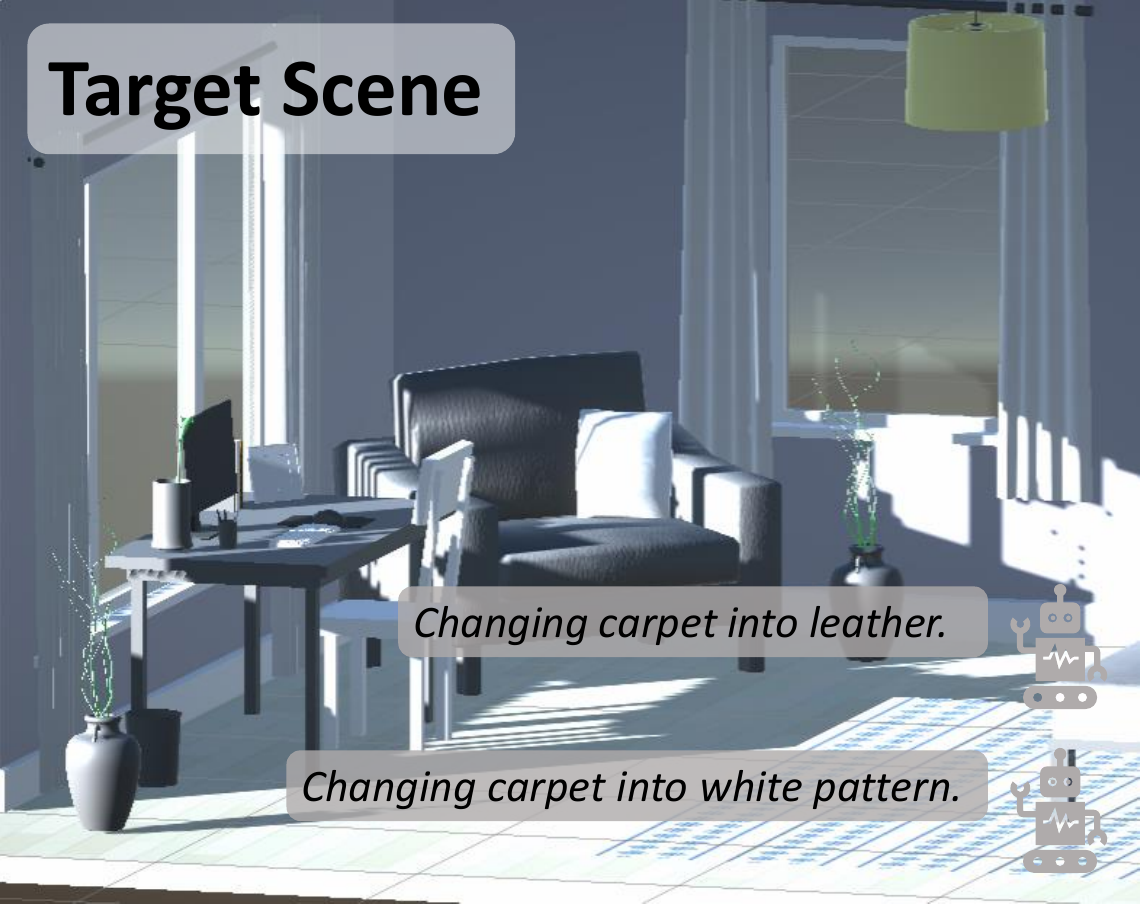}
  \caption[Example workflow for scene editing with our proposed \textsc{AssistVR} technique.]{Example workflow for scene editing with our proposed \textsc{AssistVR} technique. Left: The user adopts the \textit{bulk modification} strategy to select all blue objects in the original scene to modify their appearance together. Middle: The user adopts the \textit{incremental exploration} strategy to modify the appearance on individual objects. Right: The final scene matches the target scene of one of the tasks in our empirical user study.}
  \label{fig:teaser}
}

\abstract{
As more applications of large language models (LLMs) for 3D content in immersive environments emerge, it is crucial to study user behavior to identify interaction patterns and potential barriers to guide the future design of immersive content creation and editing systems which involve LLMs. In an empirical user study with 12 participants, we combine quantitative usage data with post-experience questionnaire feedback to reveal common interaction patterns and key barriers in LLM-assisted 3D scene editing systems. We identify opportunities for improving natural language interfaces in 3D design tools and propose design recommendations.
Through an empirical study, we demonstrate that LLM-assisted interactive systems can be used productively in immersive environments. 
} % end of abstract


\begin{document}


% \firstsection{Introduction}

\maketitle

\section{Introduction} %for journal use above \firstsection{..} instead

% \added{This text was added.}
% \deleted{This text was removed.}
% \replaced{new text}{old text}

Large Language Models (LLMs) have gained popularity in assisting task completion in immersive environments. LLMs provide various advantages to improve interaction experience in virtual and augmented reality, such as improving task completion efficiency~\cite{roberts2022steps}, democratizing VR content creation for non-expert users~\cite{giunchi2024dreamcodevr}, and improving expressiveness while reducing the user's perceived workload~\cite{kurai2024magicitem}. However, the introduction of LLMs in interaction tasks such as scene editing can also pose barriers and adversely affect the interaction experience due to the current limitations of LLMs and its capability to integrate with 3D scene content. Examples of these barriers include transparency and explainability~\cite{liao2023ai} reflected through user trust in the system, as well as appropriate error handling and timely user feedback~\cite{giunchi2024dreamcodevr}.

% To this end, Aghel Manesh et al. \cite{aghel2024people} studied prompting, 

\added{We investigate the following research questions:}

\begin{itemize}
    \item \added{\textbf{RQ1:} What common interaction patterns do users exhibit when they adopt an LLM-assisted multimodal interactive system to complete interaction tasks such as scene editing in VR?}
    \item \added{\textbf{RQ2:} Do LLM-assisted multimodal interactive systems pose interaction barriers and how do users work around them?}
\end{itemize}
\deleted{Our central hypothesis is that LLM-assisted 3D scene editing is best carried out through multimodal interaction.}
% \added{such as speech and raycast pointing}. 
\added{As speech-and-pointing interfaces have been widely studied in existing virtual and augmented reality research \cite{bolt1980put, zhou2022eliciting, zimmerer2020finally, kaiser2003mutual},}
\deleted{To begin investigating this hypothesis}
we have created the Advanced Speech Support and Interactive System for Virtual Reality (\textsc{AssistVR}), which integrates LLMs with \replaced{speech and pointing}{multimodal} interaction techniques. \added{This system serves as an example for multimodal interactive systems in general and informs our understanding of the above two research questions.} \textsc{AssistVR} uses an off-the-shelf Microsoft Azure Conversational Language Understanding (CLU) Service and GPT-4o to handle user queries. 
%\added{However, this is not the typical `LLM for X' paper and we do \textit{not} claim a system contribution. \textsc{AssistVR} is designed for the purpose of studying}
We use this system to study 
the effects of LLMs on user behavior patterns in scene editing tasks through an empirical user study with 12 participants. Specifically, we focus on whether user interaction with such LLM-assisted interactive systems reveal certain high-level \emph{interaction strategies}, and if so, was the LLM-assisted interactive system able to assist participants in identifying more efficient interaction strategies with very limited external guidance. We also examine whether the system poses any \emph{interaction barriers}, and suggest design approaches to overcome these. 
% \added{Therefore, the \textsc{AssistVR} system framework of combining Azure CLU with GPT-4o only presents one possible implementation of using LLMs to support an interaction task such as VR scene editing.}
% them as they familiarize with the system. 
Through this study, we extract observations on user performance and interaction patterns, and provide design implications for future LLM-assisted interactive systems for immersive 3D content and possibly general interactive systems which involve LLMs.
% recommendations on how LLM-assisted interactive systems can be improved to support user performance and enhance user experience.

This paper contributes to existing literature by analyzing interaction patterns and strategies through an \textit{exploratory} study. We deliberately chose not to engage in a comparative study since prior  work~\cite{de2024llmr, giunchi2024dreamcodevr, kurai2024magicitem} have proposed systems which apply LLMs to immersive content, and the capabilities of LLMs advance in a very rapid speed. Instead of making a technical contribution, this paper provides insights on observed user strategies and behavioral patterns, which generalizes to different types of interactive systems involving LLMs.

\section{Related Work}

% \subsection{Scene Understanding and Editing for Immersive 3D Content}

\subsection{Scene Editing and Multimodal Interaction in XR}

%The evolution of multimodal scene editing applications for extended reality (XR) content has been marked by a shift towards more intuitive and natural interaction paradigms. As XR technologies matured, researchers began to explore more embodied and multimodal interaction techniques.
Rakkolainen et al.~\cite{rakkolainen2021technologies} reviews recent advances in multimodal interaction technologies for extended reality (XR) content, pointing out how XR technologies introduce new interaction concepts and play an important role in addressing accessibility barriers. Similar views were proposed by Spittle et al.~\cite{spittle2022review}, who suggests that multimodal interaction facilitates selection and manipulation tasks. In 3D editing tasks in virtual reality (VR), a combined gesture and speech interface can perform on par with a unimodal interface of a radial menu in terms of promoting creativity, usability, and presence~\cite{zimmerer2020finally}.

%Within this context, various works have approached the scene editing task from different interaction modalities. In terms of gesture and voice-based editing,
Williams et al.~\cite{williams2020understanding} report on an elicitation study of speech, gesture, and multimodal speech and gesture interactions in unconstrained object manipulation tasks in augmented reality. Zhou et al.~\cite{zhou2022eliciting} found that participants preferred to use the same gesture for one and two-object manipulation in the same task, and revealed associations between speech patterns and gesture strokes during 3D object manipulation. Rodriguez et al.~\cite{rodriguez2024artists} studied natural unimodal and multimodal interaction techniques for 3D sketching in virtual reality.

%For gaze interactions,
Plopski et al.~\cite{plopski2022eye} reviewed gaze interaction and eye tracking research in XR and outlined how eye gaze has been applied to enhance user interaction with virtual content and interface design. Multimodal interactive systems such as \textsc{GazePointAR}~\cite{lee2024gazepointar} also demonstrate the possibility of leveraging eye gaze and pointing gestures to provide contextual information for speech queries.

%These works on multimodal interaction for scene editing and manipulation tasks in XR demonstrate the maturity of multimodal interaction technologies and showcase their potential in various tasks in XR. Therefore, this paper also adopts a multimodal approach to incorporate speech and raycast pointing to design a 3D editing system in virtual reality for the study on interaction strategies and patterns.

% Pros and cons of different interaction modalities? Why choose speech and raycast?

\subsection{Large Language Models for Extended Reality}

A plethora of recent research in AI and XR has focused on different aspects, including accessibility and inclusion~\cite{jiang2023beyond, bozkir2024embedding}, privacy~\cite{bozkir2024embedding}, 3D content generation~\cite{he2024enhancing, li2024discene}, and general applications~\cite{giunchi2024dreamcodevr, manfredi2023mixed, de2024llmr, song2023expanded}. Ma et al.~\cite{ma2024llms} reviews integration of LLMs with 3D spatial data as 3D-LLMs and applications. Recent work~\cite{huang2022language, huang2022inner, ahn2022can} has further explored how LLMs can assist agents in altering the physical 3D world in various ways.

% As this paper introduces the interaction patterns with an LLM-assisted system in a 3D scene editing task, we further summarize works which focus on scene editing tasks. 
In terms of 3D content editing, LLM-assisted systems such as \textsc{LLMR} \cite{de2024llmr} demonstrate a wide range of possible applications in XR, including world creation, multimodal interaction, scene editing, scene query, and integration with other external plugins, platforms, and sensors. \textsc{DreamCodeVR}~\cite{giunchi2024dreamcodevr} is an AI-powered tool for generating code in VR applications during runtime to modify the appearance and behavior of elements in a 3D scene. 
Prior work has also studied LLM prompting for immersive content. Roberts et al.~\cite{roberts2022steps} show that prompt-based methods can accelerate in-VR level editing and become an integrated part of the gameplay. Aghel Manesh et al.~\cite{aghel2024people} used a Wizard of Oz elicitation study to examine the implicit expectations of users when they prompt generative AI agents to create interactive virtual scenes.

\subsection{Interaction Pattern Analysis}

Interaction analysis is an important part of human-computer interaction (HCI) research. Wright et al.~\cite{wright2000analyzing} proposed the resources model to analyze human-computer interaction as distributed cognition, where interaction strategies play a crucial role in bringing resources in use to generate actions. 
Scholz et al.~\cite{scholz2024classifying} proposed a model to study user behavior and interaction patterns in online news forums while Guo et al.~\cite{guo2024investigating} studied interaction modes and user agency in human-LLM collaboration tasks. Beyan et al.~\cite{beyan2023co} conducted a human-human interaction analysis and identified interaction patterns and behaviors such as nonverbal cues which resulted in effective performance.
These interaction patterns are often uncovered through log analysis~\cite{trippas2024users} or audio and video analysis~\cite{jebeli2024quantifying}.

Interaction patterns have also been studied within the context of extended reality. To support the analysis of interaction patterns, symbolic event visualization methods have been proposed by Rabasahl et al.~\cite{rabsahl2023symbolic}. Feit et al.~\cite{feit2016we} and Foy et al.~\cite{foy2021understanding} studied ten-finger typing on a physical keyboard and mid-air typing in virtual reality respectively, and summarized common typing behaviors as interaction patterns. Dudley et al.~\cite{dudley2019performance} studied the performance envelopes of four alternative text input strategies in virtual reality to provide design implications for novel text entry systems.

\section{Methodology}

LLM systems have evolved from text-based interaction~\cite{gpt3} to vision-language models~\cite{tsimpoukelli2021multimodal}, which support multimodal text and images, to general-purpose multimodal LLMs~\cite{wu2023next} that support any combination of text, image, video, and audio as inputs and outputs. For immersive 3D environments, while multimodal interactive systems assisted by LLMs have been proposed~\cite{de2024llmr, konenkov2024vrgpt, giunchi2024dreamcodevr}, there is still need to investigate their effects on user behavior and interaction patterns.
We have designed \textsc{AssistVR}
\deleted{, an LLM-assisted multimodal interactive system for the purpose of studying typical interaction patterns and interaction barriers in an example task that involves editing an indoor scene to match a given target appearance using multimodal speech commands and raycast selections. The design of \textsc{AssistVR} fulfills high-level requirements of multimodal interaction and integration of LLM by incorporating speech and raycast pointing as different interaction modalities for the 3D editing task and follows a method similar to LLMR~\cite{de2024llmr} to integrate LLMs including Azure CLU and GPT-4o with scene graph information and the post-processing pipeline in Unity} to provide an integrated \replaced{speech-and-pointing 3D}{scene} editing system.

Through a scene editing user study, we gather quantitative usage data and qualitative feedback from post-experience questionnaires. The study is approved by the research ethics committee in the Department of Engineering at the University of Cambridge. Collectively, these results help us to identify main interaction strategies as well as reoccurring interaction patterns. Through post-hoc analysis of the study data, we identify key barriers in user interaction with LLM-assisted interactive systems in virtual reality and propose design implications for future LLM-assisted interactive systems.
%The remainder of this section will further introduce the methodology by providing more details on the software and hardware apparatus, participants involved in the study, the study tasks, as well as the study procedure.

% \hl{[Add stuff here]}

% Explain the overall approach and the idea of multimodal speech + pointing interaction for 3D editing in the first place --- 1--2 paragraphs

% repeat this 'we combine quantitative usage data with post-experience questionnaire feedback to reveal common interaction patterns and key barriers in LLM-assisted 3D scene editing systems'

\begin{figure*}
    \centering
    \includegraphics[width=0.65\linewidth]{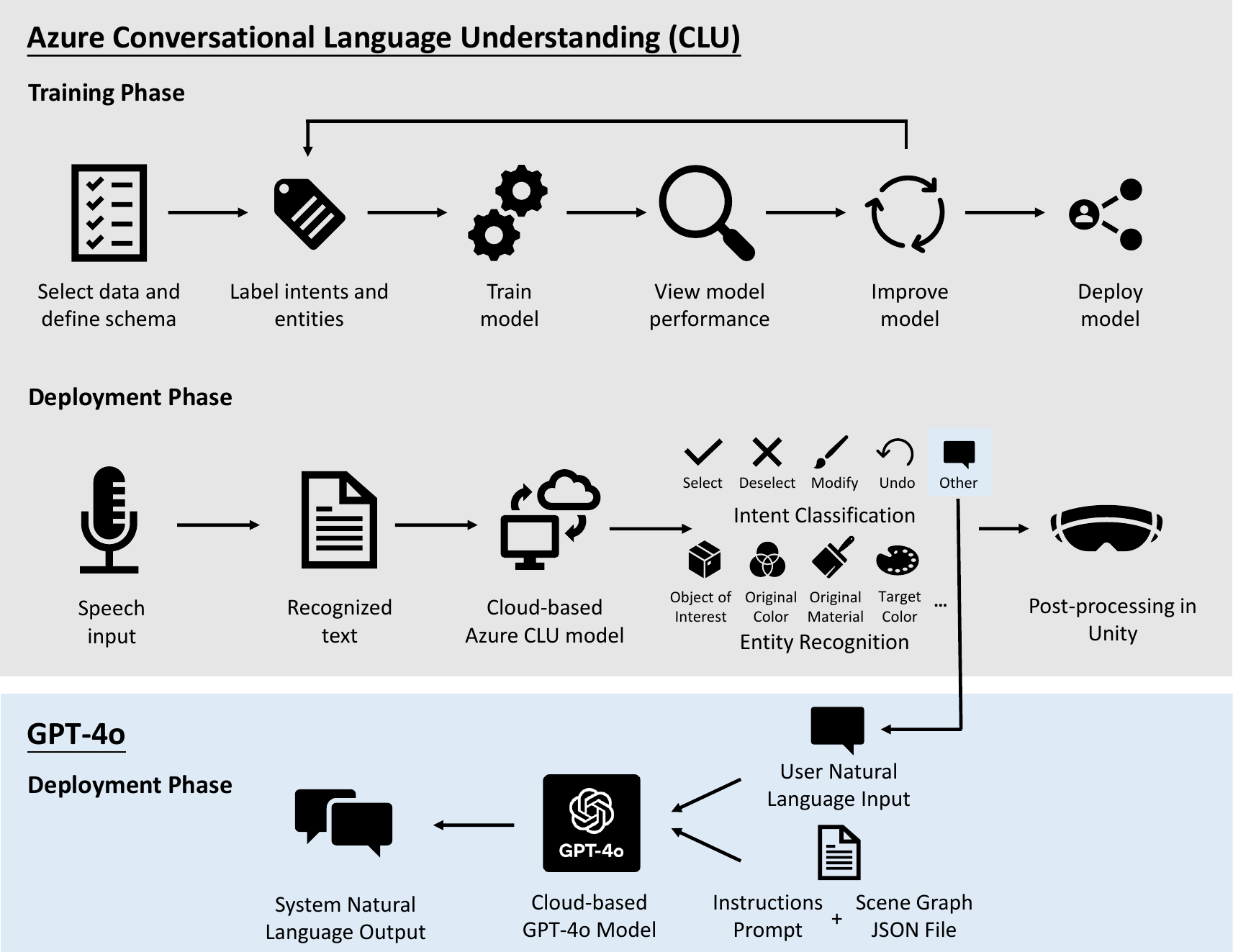}
    \caption{Workflow of the \textsc{AssistVR} system designed for the study. In the training phase, only Azure Conversational Language Understanding (CLU) is involved. The developer labels a number of utterances with intents and entities, and finetunes the Azure CLU model. The model is iteratively improved based on performance metrics. In the deployment phase, Azure CLU classifies user speech input into different intents. If the intent falls under the `Select', `Deselect', `Modify', or `Undo' categories, further post-processing steps to modify the scene are conducted in Unity. If the intent does not fall under these categories, the user speech input and a text file containing the instructions prompt and scene graph of the current scene are sent to GPT-4o, which generates a natural language response synthesized into speech for the user.}
    \label{fig:workflow}
\end{figure*}

\paragraph{Apparatus.}
To study user behavior and patterns when interacting with LLM-assisted 3D scene editing systems, we designed \textsc{AssistVR}. An outline of the system workflow is provided in \Cref{fig:workflow}. The system leverages large language models such as Microsoft Azure Conversational Language Understanding (Azure CLU, as shown in grey) and GPT-4 Omni (GPT-4o, as shown in blue)~\cite{OpenAI_2024, achiam2023gpt} to process user natural language input. \added{Azure CLU extracts intents and key entities from the user natural language input, and GPT-4o catches all exceptions which cannot be handled by the Azure CLU classifier to provide speech instructions to the user. 
}

% \added{This design provides several benefits. First, the study shows that XX\% of user commands are catched by Azure CLU, and the average delay in response time of Azure CLU is around XXX seconds. This greatly reduces the system latency and improves user experience. Only in the remaining XX\% cases when GPT-4 handles the user input, the delay is around XXX seconds due to communication with the OpenAI server and GPT-4 processing time. Second, classifying the user intent into certain predefined user intents and key entities with Azure CLU allows us to seamless integrate the system with postprocessing scripts in Unity. Compared with existing LLM-assisted interactive systems for 3D content such as LLMR \cite{de2024llmr} or DreamCodeVR \cite{giunchi2024dreamcodevr}, our approach eliminates the possibility that the LLM might generate invalid code which may break the system.
% }

In the `Training Phase' of Azure CLU, representative utterance data of possible user speech input samples are labelled with intents (such as `Select', `Deselect', `Modify', `Undo', or other intents) and key entities (such as `Object of Interest', `Original Color', `Original Material', `Target Color', and `Target Material'), and are used to finetune the default model (2022-09-01 training configuration) provided by the Azure CLU service. Upon training the model, the utterances with labeled intents and entities are adjusted to iteratively improve model performance. The final model with an F1 score of 92.73\% on intent classification was deployed. \replaced{There}{As the GPT-4o model is ready for deployment, there} is no training phase for GPT-4o.

\begin{figure}[h!]
    \centering
    \includegraphics[width=0.85\linewidth]{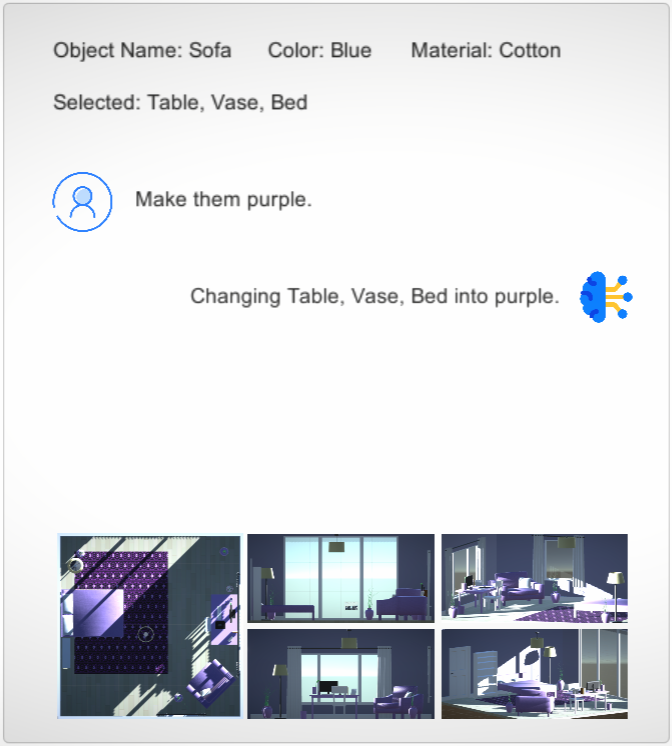}
    \caption{Example of the draggable panel. The panel shows that the current object hit by the raycast is the `Sofa' with `Blue' color and `Cotton' material. Currently, objects `Table', `Vase', and `Bed' are selected. The user says, ``Make them purple.'' The system responds, ``Changing Table, Vase, Bed into purple,'' and modifies the color of the selected objects. At the bottom of the panel, screenshots of the target scene at different angles are shown.}
    \label{fig:panel}
\end{figure}

In the `Deployment Phase' of Azure CLU and GPT-4o in \Cref{fig:workflow}, the system first uses the Azure Speech Recognition service to recognize user speech, then uses the Azure CLU model to classify the recognized speech input into different intents and extracts key entities from the user input. If the intent is classified as `Select', `Deselect', `Modify', or `Undo', the system executes post-processing scripts in Unity to perform object selection, object deselection, color/material modification, or actions to undo the previous color/material editing step. Following De La Torre et al.~\cite{de2024llmr}, the system generates a scene graph in JSON format to represent content in the 3D scene. If the intent does not fall under these four categories, the user natural language input, an instructions prompt (providing context about the scene editing task, available colors, and available materials), and a JSON file containing the scene graph of the current 3D scene are sent to the cloud-based GPT-4o model via Application Programming Interface (API) calls. GPT-4o subsequently generates a natural language response, which is then synthesized into speech and sent to the user.
% and perform selection or editing actions in VR, or synthesize natural language output to respond to the user query.  The system 
% Following De La Torre et al. \cite{de2024llmr}, the system generates a scene graph in JSON format to represent content in the 3D scene, which is fed to the GPT-4o component to generate responses to user queries. The remaining commands which can be classified as actionable intents are processed by the Azure CLU component, which evokes subsequent post-processing scripts in Unity to execute actions such as select and modify appearance in the 3D scene.

Apart from the speech-based interaction powered by Azure CLU and GPT-4o, the system also includes other interaction modalities including selection/deselection of virtual objects via raycast, and a draggable panel to provide feedback by displaying from top to bottom the name, color, and material property of the current object hit by the right raycast, the list of names of all currently-selected objects, the recognized user speech input, the natural language output from the system, as well as screenshots of the target scene at the bottom. An example screenshot of the draggable panel is provided in \Cref{fig:panel}.

Participants wore an Oculus Quest 2 headset and held the right controller during the study. The headset was connected to a Windows 10 laptop PC (Intel i5-9300H CPU, 16GB memory, and GTX 1050 graphics card) via an Oculus link cable. Scenes were implemented with Unity 3D (Version 2022.3.15f1) and publicly available resources\footnote{Source: \url{https://assetstore.unity.com/packages/3d/environments/interior-house-assets-urp-257122}.} on Unity Asset Store. 

\paragraph{Participants.} We recruited 12 participants (6 male and 6 female) aged between 22 and 35 \added{($M=26.3, SD=3.8$)}. Around 62.5\% of participants were familiar with VR, 50\% of participants were familiar with speech recognition systems, and around 67\% of participants were familiar with 3D modelling or design software. All participants understood and spoke English, and 50\% reported themselves as native English speakers. None of the participants reported any form of disability.

\paragraph{Task.} The task involves matching the original indoor scene to a target scene based on a combination of natural language instructions and image instructions. \added{Here, this scene editing task was chosen because referencing tasks and object manipulation tasks are considered as canonical interactions commonly used to evaluate interaction techniques in VR \cite{argelaguet2013survey, bergstrom2021evaluate, yu2024object}, and both tasks are encapsulated in our editing task.} The scene and task are designed such that there are a number of objects which can be referenced with a common color/material property (such as `blue' or `cotton'), and one special object (the carpet) whose target appearance can only be inspected visually by the user. The user is not instructed how to reference the pattern of the carpet verbally at the beginning of the task. The purpose of including this special object in the task is to simulate cases when objects are difficult to reference for the user and to study whether LLM-assisted systems can aid participants in referencing these objects with higher perplexity. 

Task type A involves making all blue objects in the original scene into grey and making all cotton objects in the original scene into leather, see \Cref{fig:teaser} (right) for the target scene. Task type A also involves editing the carpet into white pattern, but this requirement was given by showing images \added{instead of natural language descriptions}. \deleted{Participants were not instructed that they could use `white pattern' to refer to the target color of the carpet. }
Similarly, task type B involves making all blue objects purple, making all leather objects cotton, and making the carpet into purple pattern, see subfigure ``T: Target'' in \Cref{fig:strategy} for the target scene. The target pattern of the carpet was also given via images and not natural language. During the study, the order of task type A and task type B was counterbalanced across all 12 participants. After rearranging the order of task type A and task type B, the tasks were delivered as Task 1 and Task 2, with Task 1 preceding Task 2. 

Additional instructions for participants during Task 1 were to explore \added{the} list of all available colors and materials and to find the most efficient way to modify color and material. An additional instruction for Task 2 was to modify the scene based on the most efficient way participants found in Task 1. This reason for providing these additional instructions is because we are interested in finding whether LLM-assisted systems help participants obtain any performance improvement, and if so, how the improvement is reflected through the change in interaction strategies and patterns. \added{The complete verbal instructions for both tasks and both task types are provided in the Online Appendix.}

\paragraph{Procedure.} After filling out a consent form and demographics questionnaire, participants were briefed about the study procedure, which involved \added{asking participants to edit the color and material of various objects in a VR living room scene.}\deleted{the goals of Task 1 and Task 2 by showing them images of the original scene and target scene to match the carpet appearance and by giving them verbal instructions to make all blue objects grey (or purple) and make all cotton objects leather (or vice versa). All participants were exposed to both task type A and B in Task 1 and Task 2.} Participants were \deleted{also }told that the system supported a list of colors such as red, orange, and yellow and a list of materials such as plastic. Here, only a few examples were given, and the complete list of colors and materials were not given to the user. Users were briefed about the main functions (raycast selection, speech, and assistive panel) of the system, as well as a high-level introduction of the types of supported speech commands (Select/Deselect, Commands to modify appearance, Commands to undo, and Query commands). Participants were not taught about the exact phrases used to elicit these commands. Participants were instructed to think aloud during the study\added{. Prior to starting the two tasks, participants were given some time to familiarize themselves with the system in a practice trial. Participants were encouraged to try using speech and raycast to select and edit the color and material of a few objects, or try asking some questions on system usage and the current status of the scene.}\deleted{, and to also explore the list of all available colors and materials and find the most efficient way to modify the scene in Task 1. }

\added{After the practice trial, the goal of Task 1 was introduced to participants by showing them images of the original scene and target scene to match the carpet appearance and by giving them verbal instructions. All participants were exposed to both task type A and B in Task 1 and Task 2.}

\replaced{In Task 1, participants further explored the scene}{Participants took their time to explore the scene} and attempted the editing task. Participants also tried to figure out how to use the system efficiently, including which functions to use and what speech commands worked well. Participants were not allowed to obtain additional information from the study moderator but were allowed to ask the system. The task ended when participants were satisfied that the scene matched the target appearance. Participants were then asked to complete a post-experience questionnaire including open-ended questions, SUS~\cite{brooke1996sus}, NASA-TLX~\cite{hart1988development}, and UEQ-S~\cite{schrepp2017design} questionnaires for Task 1 and take a 5-minute break.

Next, participants were given instructions for Task 2, which included \replaced{task type A or B, as well as modifying}{making all blue objects purple (or grey), making all leather objects cotton (or vice versa), and modifying the appearance of the carpet to match the target scene. Participants were also instructed to modify} the scene based on the most efficient way they found in Task 1. After receiving the instructions for Task 2, participants modified the scene and stopped when they were satisfied that the scene matched the target appearance. Participants completed a similar post-experience questionnaire for Task 2, and gave final comments on which features they liked/disliked based on their experience throughout the entire study. The entire study lasted for around an hour. At the end of the study, participants were thanked for their participation and remunerated.

\section{Results}

Observations and quantitative data from the user study revealed several common patterns in user behavior. These findings are organized and presented below as overall performance, interaction patterns and interaction barriers. Here, significance tests do not serve to conduct comparisons between different system or interfaces, but instead serve as a tool to indicate how well users can learn to use the system over time.

\subsection{Overall Performance}\label{sec:overall-performance}

% \added{The comparison between performance metrics such as task completion quality and task completion time in Task 1 and Task 2 aims to evaluate the extent to which participants' interaction efficiency improve with an LLM-assisted 3D editing system.}\hl{remove this}

\paragraph{Task Completion Quality.} As \deleted{the task involves scene editing, and }different participants achieved different goal states which match the target scene appearance to different extents, we consider the difference between the color and material of all objects in the current scene and the color and material of all objects in the target scene as the number of \textbf{Remaining Elemental Editing Steps (REES)}, a metric to quantify user progress and task completion quality in the scene editing task. \added{Mathematically, it is defined as:
\begin{equation}
    REES = \sum_{i=1}^{n} \left[ \mathbbm{1}(o_{i(c)}\neq t_{i(c)}) + \mathbbm{1}(o_{i(m)}\neq t_{i(m)}) \right],
\end{equation}
where $\mathbbm{1}(x\neq y)$ is the indicator function. It satisfies $\mathbbm{1}(x\neq y)=1$ if $x\neq y$ and $\mathbbm{1}(x\neq y)=0$ if $x=y$. $o_i$ represents object $i$ in the scene at the current state, and $t_i$ represents the target appearance of object $i$ at the final target state. The parameter $c$ represents the color property, while $m$ represents the material property. There are in total $n$ objects in the scene.
}

% \Cref{fig:final-rees} provides a boxplot of the number of final REES of the scene when participants reported completing Task 1 and Task 2. 
This final REES metric measures how close the final state of the scene is compared to the target scene, with a lower final REES value indicating a closer match to the target scene and higher task completion quality.
\replaced{A Wilcoxon Signed-Rank test}{Friedman tests} revealed a \textbf{significant difference \replaced{($W=3, p<.05, |r|=.73$)}{($\chi^2=4.50, p<.05$)} in the final REES} between {\textsc{Task1}} ($M=4.58, SD=4.72$) and {\textsc{Task2}} ($M=1.83, SD=3.69$), suggesting that participants were able to match the target scene significantly better in Task 2 compared with their performance in Task 1. Please note, that while it can be expected that participants' performance improves over time, the scale of this improvement (60.0\% reduction in final REES on average) can indicate that users can adapt quickly to the multimodal editing system.

\begin{table*}[h!]
    \caption{\added{Participant comments organized by themes in the post-experience questionnaire.}}\label{tab:post-exp-quotes}
    \renewcommand{\arraystretch}{1.5}
    \resizebox{\textwidth}{!}{
    \begin{tabular}{p{0.075\textwidth}p{0.12\textwidth}p{0.9\textwidth}}
    \hline
    \multicolumn{1}{c}{\textbf{\added{Theme}}}                                   & \multicolumn{1}{c}{\textbf{\added{Sub-Theme}}}                           & \multicolumn{1}{c}{\textbf{\added{Participant Comments}}}   \\ \hline
    \multirow{5}{1.3cm}{\added{System Ease of Use}}     & \multicolumn{1}{l}{\added{Response efficiency}}       & \textit{\added{``It was incredibly quick. I like how efficient it was. I did not need to select anything, which made it really easy. I just told the system what to do and only had to use 5 commands.''}} \added{(P6)}   \\
                                            & \multicolumn{1}{l}{\added{Design redundancies}}                 & \textit{\added{``I very much liked the technique.   The speech recognition is much faster than selecting objects with raycast,   but I can still use the raycast to check the object properties.''}} \added{(P8)}                                                                                                                                                                                                                                                                                                                                                         \\
                                            % & \added{Straightforward to use}              & \textit{\added{``By using speech, I can easily   tell the system to select all the objects that I want to, then modify their   colors, and textures.''}} \added{(P10)}                                                                                                                                                                                                                                                                                                                                                                                                    \\
                                            & \multicolumn{1}{l}{\added{Straightforward to use}}              & \textit{\added{``It was efficient and   straightforward to use once the commands were known.''}} \added{(P11)}                                                                                                                                                                                                                                                                                                                                                                                                                                                            \\
                                            & \multicolumn{1}{l}{\added{Help and support}}    & \textit{\added{``I like how in the end I asked   the system how can I change the color \& pattern of the carpet and it   shows many examples of the exact commands that I could say. And I tried to   communicate with it and complete the task in the end and I feel like the examples   that the system gave was helpful.''}} \added{(P12)}                                                                                                                                                                                                                             \\ \hline
    \multirow{4}{1.5cm}{\added{Multimodal Interaction}} & \multicolumn{1}{l}{\added{Benefits of raycast}}                 & \textit{\added{``Raycast enabled precision   control, when I want to select, deselect a specific object that I did not   know its natural name.''}} \added{(P7)}                                                                                                                                                                                                                                                                                                                                                                                                          \\
                                            & \multicolumn{1}{l}{\added{Benefits of speech}}                  & \textit{\added{``The speech recognition is much   faster ... but I can still use the raycast   to check the object properties.''}} \added{(P8)}                                                                                                                                                                                                                                                                                                                                                                                          \\
                                            & \multicolumn{1}{l}{\added{Other possible modalities}}   & \textit{\added{``Maybe if the system is also able   to factor in my gaze or selections to provide further context. E.g., if I am   looking at a particular book and/or selected it, the voice command should   factor this in.''}} \added{(P7)}                                                                                                                                                                                                                                                                                                                           \\ \hline
    \multirow{3}{1.5cm}{\added{User Agency}}            & \multicolumn{1}{l}{\added{Unexpected response}}  & \textit{\added{``I only dislike it when I find the system not responding the way I thought it would. E.g., I selected a particular book and asked it to change book to purple, but the system changed   all the books.''}} \added{(P7)}                                                                                                                                                                                                                                                                                              \\
                                            & \multicolumn{1}{l}{\added{User adaptations}}  & \textit{\added{``Try to avoid complicating the system ... I prefer to give short and clear instructions and complete the task step by step.''}} \added{(P4)}                                                                                                                                                                                                                                                                                                                                                                                            \\ \hline
    \multirow{4}{1.2cm}{\added{User Trust}}             & \multicolumn{1}{l}{\added{Ways of establishing trust}}          & \textit{\added{``Before starting the task, I saw visually that the system was changing colors correctly for an object I had selected. I trusted it to select all objects of a certain color at once because it seemed to correctly know the color of every object. Bulk-selecting and changing seemed trustworthy and the most efficient.''}} \added{(P6)}                                                                                                                                                                                                        \\
                                            % & \added{Ways of establishing trust}          & \textit{\added{``For bulk-select, speech seems to   be faster then raycast, and the system seems to be successful to change   select and change the color for all objects. So I believed that it will do   what I say. Then I did the same for material too, but did not check if it actually   changed the materials.''}} \added{(P8)}                                                                                                                                                                                                                                   \\
                                            & \multicolumn{1}{l}{\added{Cases of a lack in trust}}            & \textit{\added{``If things are identical, I feel   confident multiselecting them using voice command. However, if things are not entirely identical, I feel better selecting and changing them one by one, so   that I don't mis-select any item that I didn't mean to.''}} \added{(P2)} \\ \hline
    \multirow{3}{2cm}{\added{Level of Feedback}}      & \multicolumn{1}{l}{\added{Visual feedback via panel}} & \textit{\added{``The panel shows what I said and what the response will be, so I can confirm if the recognition is correct or   not and try again if the recognition is wrong.''}} \added{(P8)}    \\
                                            & \multicolumn{1}{l}{\added{Other visual feedback}}      & \textit{\added{``It would be nice to view a   sample of the colours when listing them - especially for the patterns.''}} \added{(P1)}                                                                                                                                                                                                                                                                                                                                                                                                                                     \\
                                            % & \added{Other forms of visual feedback}      & \textit{\added{``I wish the speech system would   have a loading bar to visually suggest to users that if it was still   processing input information or it just cannot complete the task.''}} \added{(P12)}                                                                                                                                                                                                                                                                                                                                                              \\ 
                                            \hline
    \end{tabular}
    % \end{tabularx}
    }
\end{table*}

% People learn but the performance improvement was sustantial, fig 4 and 5 together

% \begin{figure}[h!]
%     \centering
%     \includegraphics[width=\linewidth]{figures/Final REES.pdf}
%     \caption{Box plot of final REES for Task 1 and Task 2 for all participants. Black squares indicate the mean value.}
%     \label{fig:final-rees}
% \end{figure}

\paragraph{Task Completion Time.} Another measure for task completion is the time taken for each participant to edit the scene to match the target appearance. 
% \Cref{fig:completion-time-box} provides a box plot of the task completion time for Task 1 ($M=11.2$~minutes, $SD=4.85$) and Task 2 ($M=5.74$~minutes, $SD=3.99$) for all participants. 
\replaced{A Wilcoxon Signed-Rank test}{Friedman tests} revealed a \textbf{significant difference \replaced{($W=3, p<.05, |r|=.87$)}{($\chi^2=5.33, p<.05$)} between the completion time} of \replaced{\textsc{Task1}}{Task 1} ($M=11.2$~minutes, $SD=4.9$) and \replaced{\textsc{Task2}}{Task 2} ($M=5.7$~minutes, $SD=4.0$), suggesting that participants completed Task 2 in a significantly shorter amount of time. \added{We are interested in task completion time as this metric helps to inform us on the efficiency in carrying out the tasks.
% , the significant drop in task completion time reveals a strong learning effect between both tasks.
}

% \begin{figure}[h!]
%     \centering
%     \includegraphics[width=\linewidth]{figures/completion-time-box.pdf}
%     \caption{Box plot of task completion time for Task 1 and Task 2 for all participants. Black squares indicate the mean value.}
%     \label{fig:completion-time-box}
% \end{figure}

Combining the results for task completion quality and task completion time, we observe that participants were able to match the scene significantly closer to the target scene in a significantly shorter amount of time in Task 2 after familiarizing with the system in Task 1.
% and making queries to the system to find the most efficient scene editing method. 
High standard deviations in the results also suggest that different individuals can have a vastly different performance. 
% The subsequent subsection will delve into the interaction behavior of different individuals to discuss potential patterns behind the performance of different users.

\subsection{Post-Experience Questionnaire Findings}\label{sec:post-exp-questionnaire}

% \subsection{Task Load}
Participants provided task load ratings on mental demand, physical demand, temporal demand, performance, effort, and frustration from a scale of 1 to 10 using the unweighted version of the NASA-TLX questionnaire \cite{hart1988development}.
% \Cref{fig:nasa-tlx} (left) shows a bar plot of the NASA-TLX ratings (unweighted version) \cite{hart1988development} for each category as well as the overall load with 95\% confidence intervals of the mean score.
A Wilcoxon signed rank test revealed that the \textbf{overall task load rating of {\textsc{Task1}} ($M=3.97, SD=1.08$) was significantly higher ($W=4, p<.05, |r|=.78$) than that of \textsc{Task2} ($M=3.35, SD=.98$)}. 
% Results are summarized in \Cref{tab:QuestionnaireResults}.

% \Cref{fig:nasa-tlx} (right) presents a bar plot of the 
Results from the System Usability Scale (SUS) \cite{brooke1996sus} of \textsc{Task1} and \textsc{Task2} yielded a higher average SUS score in Task 2 compared with Task 1.
% with 95\% confidence intervals of the mean estimate. 
However, a Wilcoxon signed rank test did not reveal a significant difference \replaced{($W=9.5, p=.073, |r|=.57$)}{($W=9.5, p=.073, r=.568$)} between the SUS ratings of \textsc{Task1} ($M=72.1, SD=15.5$) and \textsc{Task2} ($M=75.2, SD=14.9$).
% Results are summarized in \Cref{tab:QuestionnaireResults}.

Results from the short version User Experience Questionnaire (UEQ-S) \cite{schrepp2017design} show that \textsc{Task2} attains a higher average overall score ($M=1.70, SD=.81$) compared with \textsc{Task1} ($M=1.50, SD=.59$), but Wilcoxon signed rank tests did not reveal a significant difference ($W=13, p=.083, |r|=.52$). For the subcategories of the UEQ-S ratings, no significant differences were found in the \textsc{Pragmatic} quality ($W=14.5, p=.199, |r|=.41$) between \textsc{Task1} ($M=1.50, SD=.80$) and \textsc{Task2} ($M=1.79, SD=1.09$) or the \textsc{Hedonic} quality ($W=0, p=.089, |r|=1.0$) between \textsc{Task1} ($M=1.50, SD=1.06$) and \textsc{Task2} ($M=1.60, SD=1.04$).
% \Cref{fig:ueq-s} shows the 
% Results from the short version User Experience Questionnaire (UEQ-S) \cite{schrepp2017design} show that \textsc{Task1} attains a higher average pragmatic quality score
% % and a higher average overall scale, 
% and \textsc{Task2} attains a higher average hedonic quality score. 
% Wilcoxon signed rank tests reveal a \textbf{significant difference ($W=6, p<.05, |r|=.737$) in the overall UEQ-S score between \textsc{Task1} ($M=1.50, SD=.590$) and \textsc{Task2} ($M=1.22, SD=.640$)}. For the subcategories of the UEQ-S ratings, \textbf{significant differences were found in the \textsc{Pragmatic} quality ($W=6, p<.05, |r|=.738$) between \textsc{Task1} ($M=1.50, SD=.798$) and \textsc{Task2} ($M=.833, SD=.587$)}.
% Results are summarized in \Cref{tab:QuestionnaireResults}.

% \hl{[Open-ended question comments?]}
\deleted{Following the questionnaires on task load, system usability, and user experience, participants}\added{Participants} also provided descriptions of the most efficient strategy they found, as well as open comments about the system.
% Most efficient strategy
Ten out of twelve participants were able to find an efficient strategy of bulk-editing object properties by interacting with the system without additional external assistance by the end of the study.
Participants who did not find the bulk modification strategy described their strategy as follows\added{.}\deleted{:} \added{P2 said, ``For identical items such as blue walls, blue vases and leather pillows, I tend to use voice command to change their colours/material...For non-repetitive items such as the pen holder and keyboard, I just selected and changed them individually one by one.'' Meanwhile, P12 commented, ``Because I found selecting multiple objects at the same time [being] troublesome, I directly ask[ed] the speech system to help chang[e] the color of multiple objects.''} 
% \hl{add deleted quote here}
% \begin{quote}
%     \textit{\deleted{``For identical items such as blue walls, blue vases and leather pillows, I tend to use voice command to change their colours/material...For non-repetitive items such as the pen holder and keyboard, I just selected and changed them individually one by one.''}} \deleted{(P2)}
% \end{quote}
% \begin{quote}
%     \textit{\deleted{``Because I found selecting multiple objects at the same time [being] troublesome, I directly ask[ed] the speech system to help chang[e] the color of multiple objects...I changed 2-3 items together just to make the system work effectively.''}} \deleted{(P12)}
% \end{quote}
These two participants either found it more reassuring to change individual objects (P2) or did not find the supported command or workflow to select multiple objects with the same property first and then use another command to modify the appearance of all selected objects (P12). \replaced{A thematic}{Further} analysis \cite{guest2012introduction} on open comments about the system revealed the following trends, which provide further insights on the observed behaviors and interaction patterns\replaced{.}{:} \added{For the following themes, examples of quotes from individual participants are provided in \Cref{tab:post-exp-quotes}, while participant conversation histories are provided in the Online Appendix.}

\paragraph{System ease of use.} 
Participants appreciated how easy and efficient it was to complete the scene editing task once they knew how to phrase the commands and which strategy to adopt. \deleted{P6 found the system to be \textit{``incredibly quick''}, and liked how efficient the system was. P6 commented, ``I did not need to select anything, which made it really easy. I just told the system what to do and only had to use 5 commands.'' P8 liked how the system supported multiple ways to complete the task. P10 commented that with speech commands, he could easily tell the system to select objects, then modify their colors and textures. P11 commented, ``It was efficient and straightforward to use once the commands were known.''}
% \begin{quote}
%     \textit{``I like how in the end I asked the system how can I change the color \& pattern of the carpet and it shows many examples of the exact commands that I could say.''} (P12)
% \end{quote}
Dialogues between participants and \textsc{AssistVR} \deleted{during the study }revealed that participants found the system useful in providing suggestions to help them find the list of all supported colors, materials, supported speech commands, as well as the efficient bulk modification strategy.  \deleted{For example, P12 asked, ``Can you make what I selected into another material,'' and the system responded, ``Yes, you can change the material of the selected objects. Simply use a speech command like `Make it leather' or `Make it cotton' to modify the material of the selected objects.'' P2 asked, ``What are the options for colours,'' and the system responded, ``You can change the color of an object to red, orange, yellow, ..., yellow pattern, purple pattern, or white pattern.'' P6 asked, ``What materials can I use,'' and the system responded, ``You can change the material of an object to plastic, cotton, leather, or concrete.'' P5 asked, ``What is the fastest way to change all the blue objects to purple,'' and the system responded, ``The fastest way to change all the blue objects to purple is to use speech commands. First, press Button A on the right controller and say 'Select all blue objects'. After selecting all blue objects, use the speech command 'Make them purple'.'' In the post-experience questionnaire, P12 also expressed how she liked how the system `shows many examples of the exact commands that [she] could say' when asked how to change the color and pattern of the carpet.}

\paragraph{Multimodal Interaction.} In the post-experience questionnaire, participants appreciated how different interaction modalities including speech and raycast worked together to facilitate scene editing tasks. \deleted{P7 commented that raycast interaction enabled `precision control' when users do not know the precise name of objects, while speech interaction is helpful when there is little uncertainty around the object name or command. P8 `very much liked the technique' because the speech modality allowed her to select objects in a fast manner, while the raycast modality helped her check object properties easily. Comments from other participants also show how multimodal interaction techniques can be helpful in LLM-assisted interactive systems.}
% \hl{add deleted quote here}
% \begin{quote}
%     \textit{\deleted{``I like how intuitive both (speech and raycast) were, this made me more comfortable using them.''}} \deleted{ (P11)}
% \end{quote}
% \begin{quote}
%     \textit{\deleted{``Speech was easiest for this task, but raycast was also useful for selecting the carpet.''}} \deleted{ (P1)}
% \end{quote}
% \begin{quote}
%     \textit{\deleted{``I like the raycast for selecting individual object as it's more accurate, while prefer the speech for selecting multiple identical objects as it's quicker.''}} \deleted{(P2)}
% \end{quote}
\replaced{Some participant comments also demonstrated that multimodal interaction is not limited to speech and raycast but can instead incorporate a broader range of interaction modalities in future systems.}{
When asked which features did participants wish to have, P7 further commented, ``Maybe if the system is also able to factor in my gaze or selections to provide further context,'' which further demonstrates that multimodal interaction is not limited to speech and raycast but can instead incorporate a broader range of interaction modalities in future systems.}
% \begin{quote}
%     ``Maybe if the system is also able to factor in my gaze or selections to provide further context.'' (P7)
% \end{quote}

\paragraph{User Agency.} Participants commented how sometimes the system did not respond to \deleted{user }speech input as they expected, which negatively affected their sense of agency over the system. \deleted{P7 commented that he disliked it when the system did not respond as he expected, for example when the system selected all books when he referred to the singular `book' in his command.} During the study, participants were aware of gaining control of their actions and sought to improve \deleted{user }agency by choosing appropriate \deleted{interaction }strategies. \deleted{For example, P4 preferred to `select all relevant items' and `change one property a time, color first and then material,' because she preferred to `give short and clear instructions and complete the task step by step' and `avoid complicating the system'.}
% This decrease in user agency can potentially influence user strategy. When asked about the reasons for choosing their interaction strategy, P4 commented that she tried to avoid complicating the system 

% \begin{quote}
%     \textit{``I only dislike it when I find the system not responding the way I thought it would. E.g., I selected a particular book and asked it to change book to purple, but the system changed all the books.''} (P7)
% \end{quote}

% \begin{quote}
%     ``Select all relevant items; Change one property a time, color first and then material.'' (P4)
% \end{quote}

% \begin{quote}
%     ``Try to avoid complicating the system or multitasking. I prefer to give short and clear instructions and complete the task step by step.'' (P4)
% \end{quote}

\paragraph{User Trust.} Post-experience comments revealed that some participants sought simple ways to verify that the system was processing speech commands correctly before bulk-editing the scene. \deleted{P6 commented that before starting the task, she saw visually that the system was changing colors correctly for selected objects, and she `trusted it to select all objects of a certain color at once because it seemed to correctly know the color of every object'. P6 and P8 confessed that they simply trusted the system to do everything correctly upon simple verification that the system seems to be successful in selecting objects and changing their colors, and they would not know if the system made a few mistakes. On the other hand, P2 did not completely trust the system and felt `better selecting and changing [objects] one by one' if `things are not entirely identical', such that he does not `mis-select any item' and `a larger margin for error can be ensured'.} The difference in user trust in the system also likely led them to choose different interaction strategies. \deleted{P6 and P8 who trusted the system used \textit{bulk modification} in both Task 1 and Task 2, whereas P2 who did not trust the system did not try \textit{bulk modification} in either Task 1 or Task 2.}

% \begin{quote}
%     ``If things are identical, I feel confident multiselecting them using voice command. However, if things are not entirly identical, I feel better selecting and changing them one by one, so that I don't mis-select any item that I didn't mean to, and so that a larger margin for error can be ensured.'' (P2)
% \end{quote}

% \begin{quote}
%     ``Before starting the task, I saw visually that the system was changing colors correctly for an object I had selected. I trusted it to select all objects of a certain color at once because it seemed to correctly know the color of every object. Bulk-selecting and changing seemed trustworthy and the most efficient. '' (P6)
% \end{quote}

% \begin{quote}
%      ``I suppose I just trusted the system to do everything correctly. If the system did something wrong, I probably would not have known.'' (P6)
% \end{quote}

% \begin{quote}
%     ``For bulk-select, speech seems to be faster then raycast, and the system seems to be successful to change select and change the color for all objects. So I believed that it will do what I say. Then I did the same for material too, but did not check if it actually changed the materials.'' (P8)
% \end{quote}

\paragraph{Level of feedback.} Participants appreciated how the system provided an adequate amount of visual feedback via the draggable panel and voice feedback through synthesized speech (see quotes under `System ease of use'). 
% \hl{add deleted quote here}
% \begin{quote}
%     \textit{\deleted{``The panel shows what I said and what the response will be, so I can confirm if the recognition is correct or not and try again if the recognition is wrong.''}} \deleted{(P8)}
% \end{quote}
Participants also commented how it would be helpful if they received more visual feedback on the list of colors and materials, in addition to their names. \deleted{P1 commented that it would be helpful to `see a sample of the colours', `especially for the patterns', rather than listing only the names. P12 also suggested providing visual feedback in the form of a progress bar to indicate that the LLM is still processing information, or whether it is unable to complete the task.}
% \begin{quote}
%     ``Sometimes the speech recognition doesn't do what I wanted it to complete. It requires a specific way of asking.'' (P8)
% \end{quote}
% \begin{quote}
%     \textit{``Would be helpful to see a sample of the colours, rather than just listing the names.''} (P1)
% \end{quote}
% \begin{quote}
%     ``It would be nice to view a sample of the colours when listing them - especially for the patterns.'' (P1)
% \end{quote}

\subsection{Interaction Patterns}\label{sec:interaction-patterns}

\begin{figure}[h!]
    \centering
    \includegraphics[width=\linewidth]{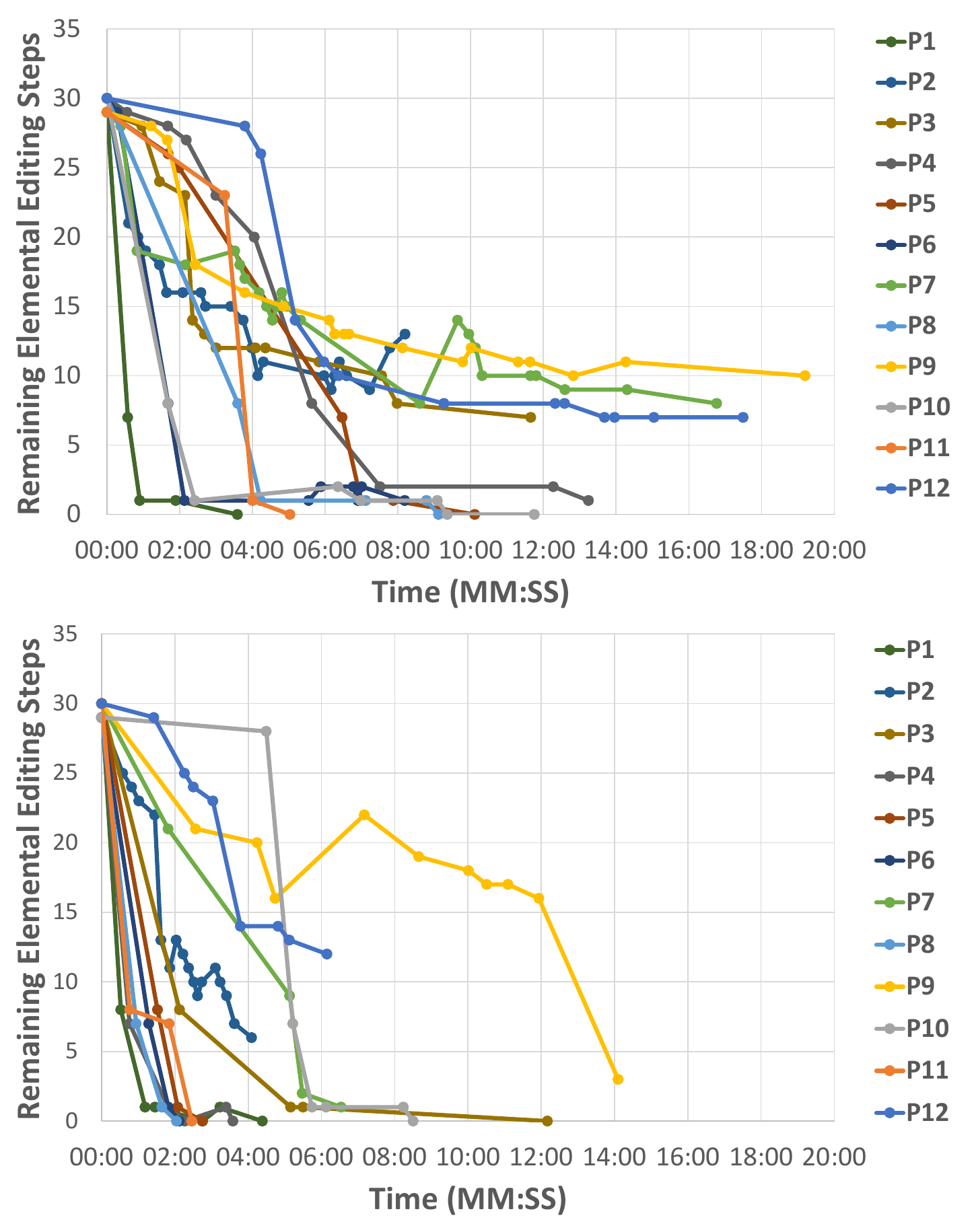}
    \caption{Number of remaining elemental editing steps to match the target scene in Task 1 (top figure) and Task 2 (bottom figure). The horizontal axis is denoting relative time in minutes and seconds.
    % Some participants preferred to edit the scene iteratively (grey region), while others preferred to bulk edit the appearance of multiple objects (green region).
    }
    \label{fig:remaining-steps}
\end{figure}

The study revealed how participants preferred to iteratively modify the color and material of individual objects in the scene to match the target appearance, or to select a group of objects with a common feature, and change their color/material using a single voice command.  \Cref{fig:remaining-steps} plots the number of remaining elemental editing steps for all 12 participants with respect to elapsed time. 

%  Figure 6 span 2 columns, increase font

As shown in \Cref{fig:remaining-steps}, in Task 1,  in which participants are asked to find the most efficient way to edit the scene to match the target appearance, P2, P3, P7, P9, and P12 preferred to make incremental edits to individual objects. Similarly in Task 2, in which participants are asked to edit the scene based on the most efficient method they found, P2, P9 and P12 also preferred to modify the scene iteratively. We define this high-level scene editing strategy as:
\begin{description}
\item [Incremental Exploration (IE)] This strategy emphasizes visual inspection of individual object properties and combines raycast selection or speech selection of single objects by their names and modifying object appearance using speech commands, or direct modification (without explicit selection) of individual object appearance through speech commands.
\end{description}

%  increase font in fig 7. separate 

\begin{figure*}[h!]
    \centering
    \includegraphics[width=0.495\linewidth]{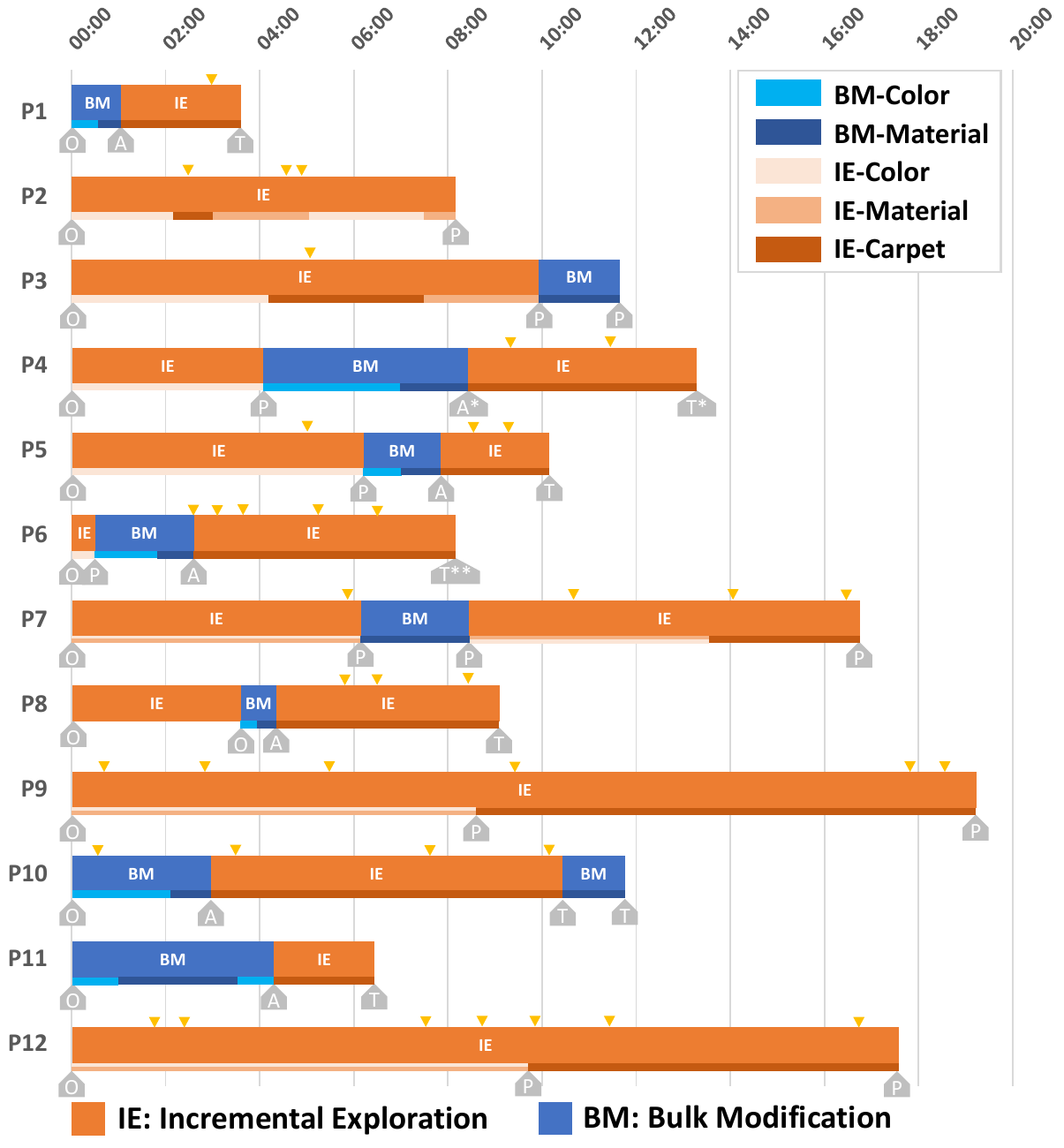}
    \includegraphics[width=0.495\linewidth]{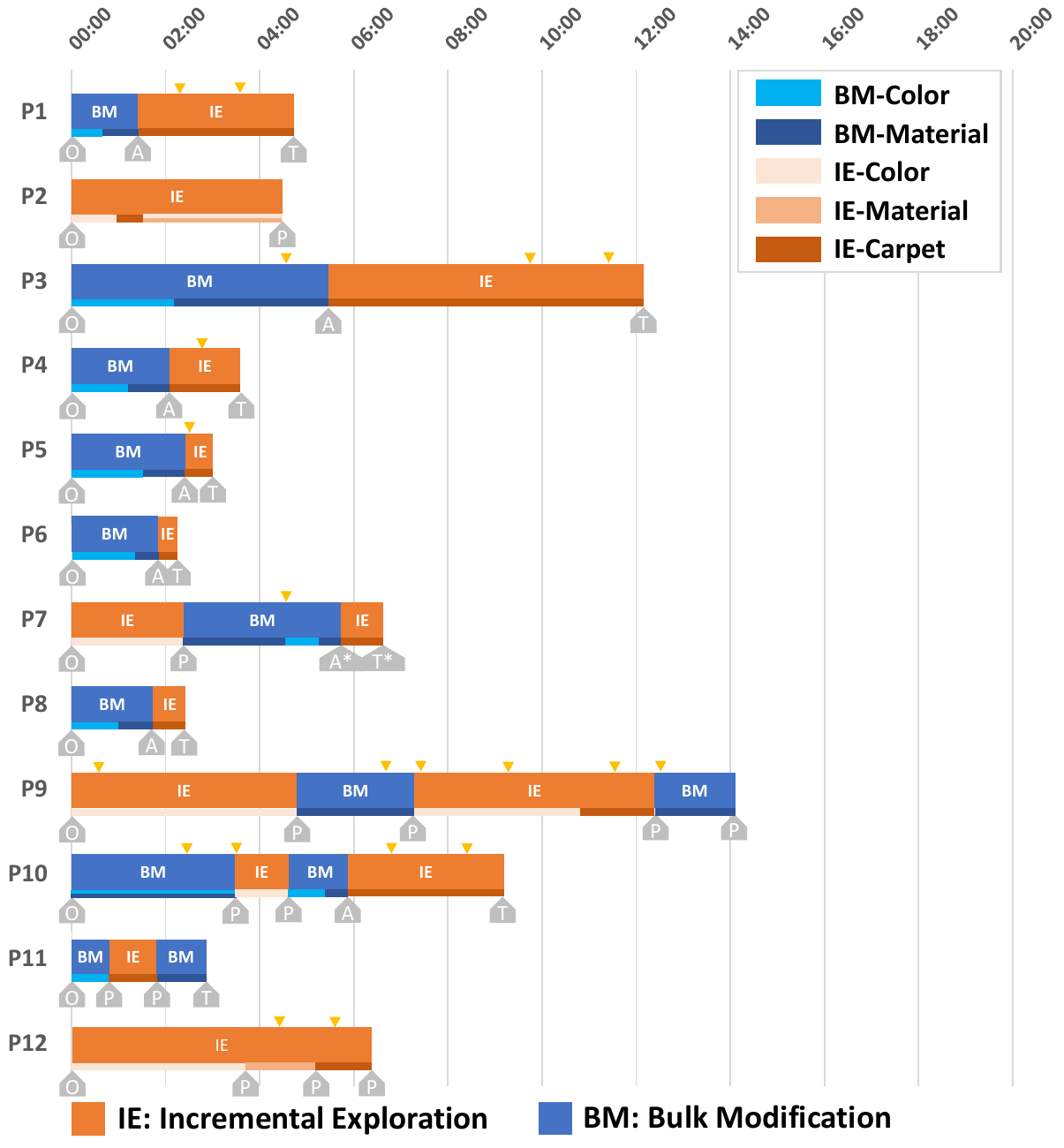}
    
    \vspace{1em}
    
    \includegraphics[width=0.248\linewidth, page=1]{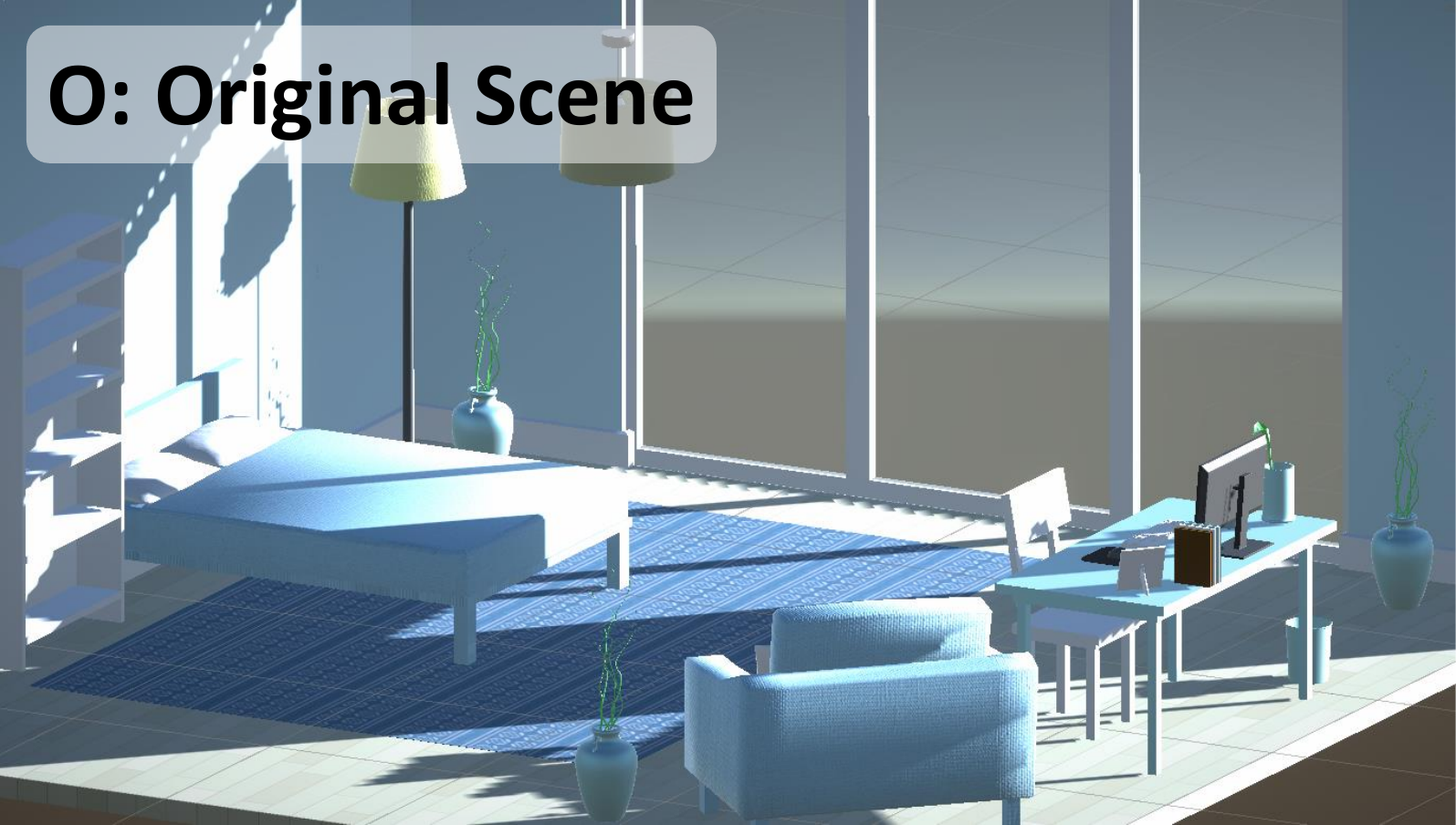}
    \includegraphics[width=0.248\linewidth, page=2]{figures/scene-screenshots.pdf}
    \includegraphics[width=0.248\linewidth, page=3]{figures/scene-screenshots.pdf}
    \includegraphics[width=0.245\linewidth, page=4]{figures/scene-screenshots.pdf}
    \includegraphics[width=0.245\linewidth, page=5]{figures/scene-screenshots.pdf}
    \includegraphics[width=0.245\linewidth, page=6]{figures/scene-screenshots.pdf}
    \includegraphics[width=0.245\linewidth, page=7]{figures/scene-screenshots.pdf}
    \vspace{-0.5em}
    \caption{Interaction strategies adopted by different users across the duration of Task 1 (top left) and Task 2 (top right). Time on the horizontal axis is displayed in the format MM:SS (minutes:seconds). Triangles above the timeline of each participant indicate user queries. The main timeline bar for each participant indicates the high-level strategy employed (IE: Incremental Exploration, or BM: Bulk Modification). The secondary timeline bar below the main timeline bar indicates the low-level strategy employed, namely Bulk Modify Color (BM-Color), Bulk Modify Material (BM-Material), Color Editing with Incremental Exploration (IE-Color), Material Editing with Incremental Exploration (IE-Material), or Carpet Editing with Incremental Exploration (IE-Carpet). Grey tags below the timeline bars represent the current scene status (O: Original scene. T: Target scene. P: Partially-edited scene.) The target scene can be the grey scene in \Cref{fig:teaser} (right), or the purple scene shown here. \deleted{Each participant experienced both target scenes, and the order of the target scene in Task 1 and Task 2 is counterbalanced for all participants. }Among partially-edited scenes, some scenes occur frequently and are labelled explicitly. These include: T*: In addition to T, one of the walls received an extra edit, REES=1. T**: In addition to T, the carpet material is incorrect, REES=1. A: Color/material changed for all objects except the carpet, REES=1. A*: In addition to A, one of the walls received an extra edit, REES=2. Part of the secondary timeline for P8 in Task 1 is blank because only selections instead of edits occurred. \replaced{Examples}{Example screenshots} of these scenes are provided below the timeline.}
    \label{fig:strategy}
    \vspace{-1.1em}
\end{figure*}

In Task 1, P1, P6, P8, P10, P11 used speech commands to select a group of objects via their common color or material property and used a single voice command to bulk edit their appearance. This strategy is also observed in Task 2 within the behavior of more participants including P1, P3, P4, P5, P6, P7, P8, P10, and P11. We define this high-level interaction strategy as:

\begin{description}
% \item [Incremental Exploration] This strategy emphasizes visual inspection of individual object properties and combines raycast selection or speech selection of single objects by their names and modifying object appearance using speech commands.
% In this strategy users tended to ...
\item[Bulk Modification (BM)] This strategy uses speech to select a group of objects with a shared color/material property, then uses speech to bulk modify their appearance. In this strategy, there is not explicit involvement of visual inspection of individual object properties.
\end{description}

It is important to note that the interaction strategy of a certain user can change over time. For example in Task 1, P4 started the task with \textit{incremental exploration}, then adopted \textit{bulk modification}, before returning to \deleted{the }\textit{incremental exploration}\deleted{ strategy}. Therefore, we visualize how interaction strategies have changed (if any) over the course of time in Task 1 and Task 2 for each participant in \Cref{fig:strategy}.
Based on these interaction patterns, we make the following observations:

\paragraph{Color modification tends to precede material modification in IE and BM.} In Task 1, among all 12 participants, 7 edited color before editing material (P1, P3-P6, P8, P10) while none edited material before editing color. The remaining 5 participants did not exhibit a strong preference on editing a certain property before another (P2, P7, P9, P11, P12). In Task 2, 8 participants edited color before editing material (P1, P3-P6, P8, P11, P12) while none edited material before editing color. The remaining 4 participants did not exhibit a strong preference on the editing sequence (P2, P7, P9, P10). This trend in editing sequence regardless of the high-level strategy employed demonstrates how the majority of participants drew attention to the more distinguishable visual features such as object colors and edited these features first in comparison with less distinguishable visual features such as object material.

\paragraph{Carpet tends to be edited last.} Prior to the study, participants were instructed to match the appearance of the carpet to the appearance shown in an image of the target scene. Participants were not explicitly told how to modify the carpet appearance, or how to reference the target appearance of the carpet. In comparison, the remaining objects were given an explicit target color (grey or purple). The carpet represents objects which are difficult to edit verbally, and the study results revealed that in Task 1, 9 out of 12 participants (P1, P4--P9, P11, P12) chose to edit the carpet last. In Task 2, 9 out of 12 participants (P1, P3--P8, P10, P12) edited the carpet after editing the remaining objects. The results show that in speech-based interfaces, users tend to edit objects with a clear goal state such that the speech commands are easy to enunciate.
% \hl{[Could be because the instruction to modify the carpet was given last. put in limitation.]}

\begin{figure*}[t]
    \centering
    \includegraphics[height=4.8cm]{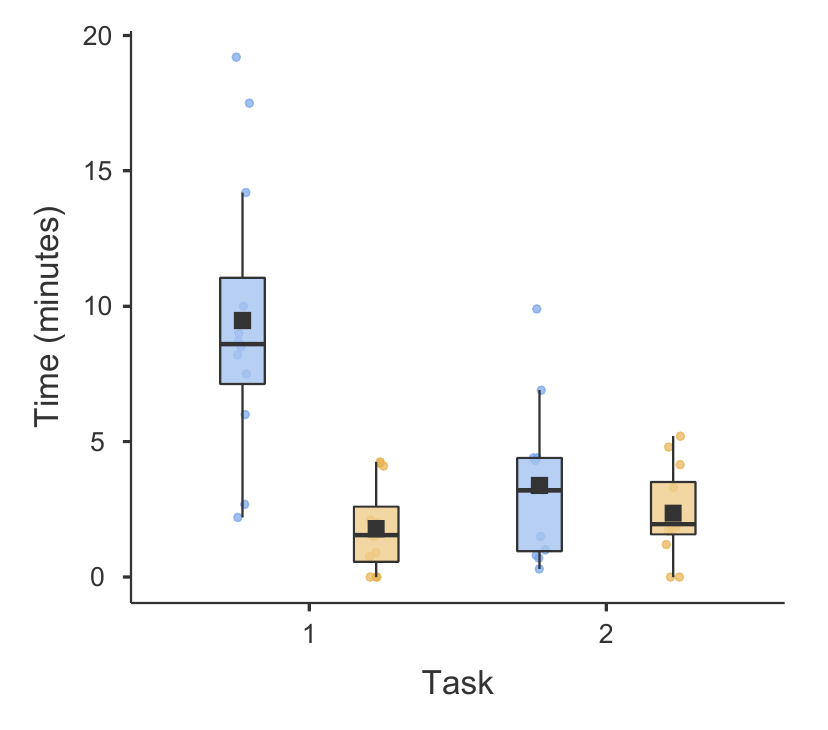}
    \includegraphics[height=4.8cm]{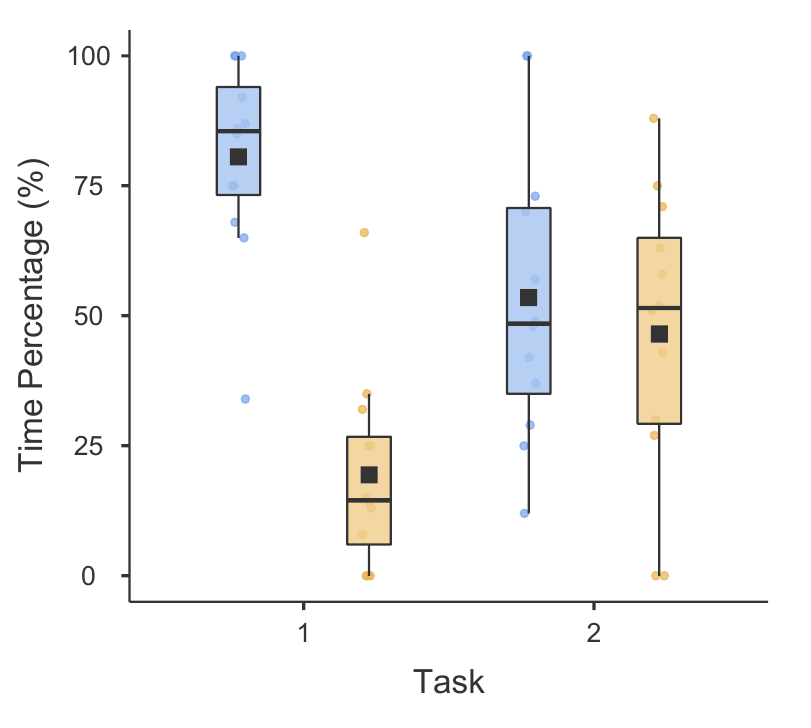}
    \includegraphics[height=4.8cm]{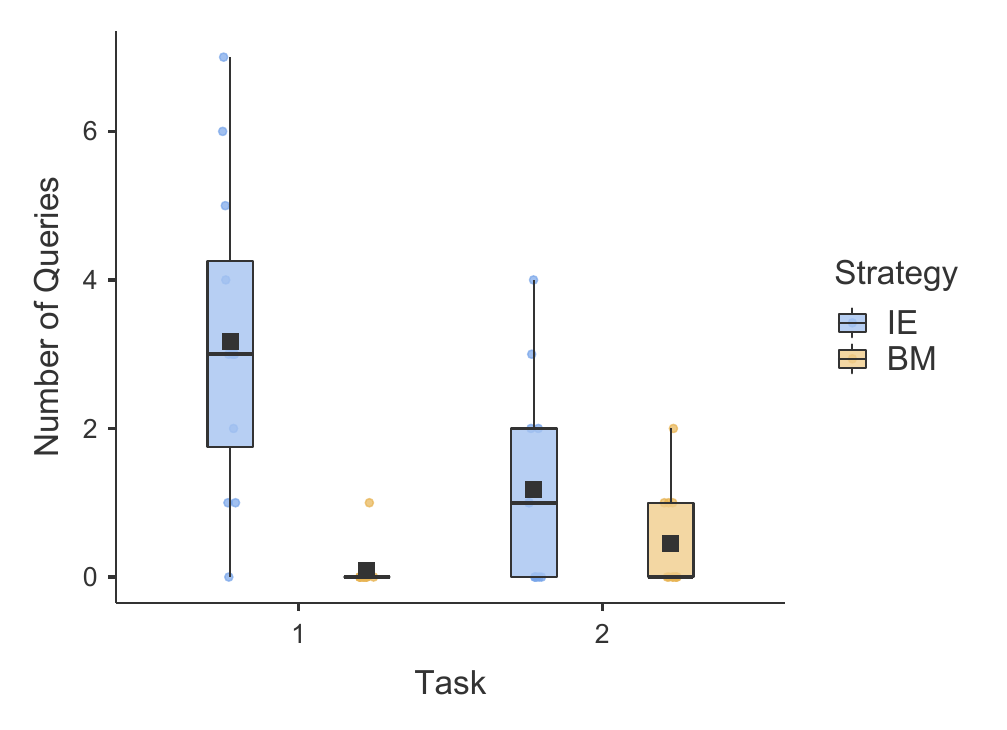}
    \vspace{-1em}
    \caption{Box plots of the time spent in minutes for all participants on the incremental exploration (IE) strategy and the bulk modification (BM) strategy (left), the percentage of time spent on both strategies for all participants in Task 1 and Task 2 (middle), and the number of queries posed during both strategies for all participants in Task 1 and Task 2 (right). Black squares indicate the mean values.}
    \label{fig:strategy-time-box}
    \vspace{-1em}
\end{figure*}

\paragraph{Total time spent on incremental exploration tends to exceed the time spent on bulk modification.} \Cref{fig:strategy-time-box} (left) provides a box plot of the time spent on incremental exploration and the time spent on bulk modification for all 12 participants. \replaced{Wilcoxon Signed-Rank tests}{Friedman tests} indicate that \textbf{the total minutes spent on the incremental exploration strategy ($M=9.47, SD=5.22$) in Task 1 is significantly greater \replaced{($W=2, p<.05, |r|=.92$)}{($\chi^2=8.33, p<.05$)} than the minutes spent on the bulk modification strategy ($M=1.78, SD=1.62$)}. For Task 2 however, the difference between the time taken on incremental exploration ($M=3.38, SD=2.87$) and bulk modification ($M=2.36, SD=1.70$) was not significantly different \replaced{($W=28, p=.424, |r|=.23$)}{($\chi^2=.33, p=.564$)}.

\paragraph{A larger percentage of time was spent on Bulk Modification in Task 2 compared to Task 1.} \Cref{fig:strategy-time-box} (middle) provides box plots of the percentage of time spent on incremental exploration and bulk modification for each participant. \replaced{Wilcoxon Signed-Rank tests}{Friedman tests} indicate that \textbf{the percentage of time spent on bulk modification for each participant significantly increased \replaced{($W=2, p<.05, |r|=.81$)}{($\chi^2=6.40, p<.05$)} in Task 2 ($M=.465, SD=.279$) compared with Task 1 ($M=.194, SD=.190$)}. This suggests that users are likely to have learned about the efficiency of the bulk modification strategy and prefer to spend more time on it.

% \begin{figure}[h!]
%     \centering
%     \includegraphics[width=\linewidth]{figures/strategy-time-box-percent.pdf}
%     \caption{Box plots of the percentage of time spent on the incremental exploration (IE) strategy and the bulk modification (BM) strategy for all participants in Task 1 and Task 2. Black squares indicate the mean values.}
%     \label{fig:strategy-time-box-percent}
% \end{figure}

\paragraph{More queries were posed during IE.}
\Cref{fig:strategy-time-box} (right) provides box plots of the number of queries posed during incremental exploration ($M=3.17, SD=2.08$) and bulk modification ($M=.08, SD=.29$) for all participants in Task 1 and the number of queries posed during incremental exploration ($M=1.25, SD=1.36$) and bulk modification ($M=.42, SD=.67$) for all participants in Task 2. \replaced{Wilcoxon Signed-Rank tests}{Friedman tests} indicate that \textbf{significantly more queries were posed during incremental exploration, as compared to bulk modification in both Task 1 \replaced{($W=0, p<.05, |r|=.87$)}{($\chi^2=11.0, p<.001$)} and Task 2 \replaced{($W=2.5, p<.05, |r|=.76$)}{($\chi^2=4.50, p<.05$)}}.

% \begin{figure}[h!]
%     \centering
%     \includegraphics[width=\linewidth]{figures/num-queries-box.pdf}
%     \caption{Box plots of the number of queries posed during incremental exploration (IE) and bulk modification (BM) for all participants in Task 1 and Task 2. Black squares indicate the mean values.}
%     \label{fig:num-queries-strategy-box}
% \end{figure}

% \paragraph{More queries were posed in Task 1.} Following the previous observation, on average more queries were posed in Task 1 compared with Task 2 for both the incremental exploration strategy ($\chi^2=3.60, p=.058$) and the bulk modification strategy ($\chi^2=3.00, p=.083$).

\paragraph{Queries did not necessarily guide participants to find the BM strategy.} P2, P9, and P12 who did not try the bulk modification strategy in Task 1 posed 3, 6, and 7 queries respectively, but only P9 shifted to a combination of the incremental exploration strategy and bulk modification strategy in Task 2, while P2 and P12 continued with the incremental exploration strategy and were not successful in matching the scene to the target appearance.

\paragraph{Participants who tried BM in Task 1 achieved high performance in Task 2.} P1, \replaced{P3 to P8}{P3, P4, P5, P6, P7, P8}, P10, and P11 tried the bulk modification strategy in Task 1. All \deleted{these }participants were able to complete Task 2 to match exactly the target scene (P1, P3-P6, P8, P10, P11) or match sufficiently close (T*) to the target scene (P7). \replaced{Wilcoxon Signed-Rank tests}{Friedman tests} on the task completion time of these participants also revealed a \textbf{significantly shorter \replaced{($W=3, p<.05, |r|=.78$)}{($\chi^2=2.78, p<.05$)} completion time of Task 2 compared with Task 1}.

% \paragraph{Tendancy to start with IE first in Task 1.}

% \paragraph{Number of raycast selections.}

% \paragraph{Relationship between successful queries and performance in subtasks (color, material, carpet).}

\subsection{Interaction Barriers}

The study also revealed certain interaction barriers which adversely affected the completion quality of scene editing tasks or the completion efficiency of the tasks.

\paragraph{Speech Recognition/Processing Issues.}

The misrecognition and processing errors of certain words by the system required participants to repeat their queries multiple times, which ultimately resulted in a delay in task completion. Interaction barriers under this category can be due to a recognition error from the Microsoft Azure Speech Recognition service, or due to a processing error in the misrecognition of user intents or key entities by Azure CLU or an error occurred when matching key entities to the GameObjects or textures in Unity during the Unity post-processing step.
% The system sometimes doesn't recognize subtle differences in commands (e.g., changing color vs. changing pattern of the carpet).

\paragraph{Feedback Clarity.}

In the user study, participants were often confused when the system failed to respond to user speech input according to user expectations. In such circumstances, the system does not always provide clear feedback on why a command did not work or instruct participants on how to phrase it correctly. For queries that were not categorized as `Other', the Azure CLU model did not have the capability to provide feedback to users. For speech input processed by GPT-4o, the model lacked enough contextual information on the current status of each object (such as whether they are selected or not) in the scene and could not provide enough feedback. Participants also commented how visual feedback could be further improved by, for example,  adding images to describe colors and materials in addition to text, or adding a progress bar to indicate that the LLM is processing the user query. 

\paragraph{Command Phrasing.}

Some participants had difficulty finding the correct phrasing for some commands, especially for changing patterns or materials. This is because the training data of the Azure CLU model only labelled commands with a certain sentence structure as selection or editing commands. Commands with different phrasing are processed by GPT-4o, but the model often replied that it did not have the capability to select or modify objects, which resulted in confusion among participants. This example shows how special considerations should be included in the GPT prompt to instruct LLMs to incorporate information from other sources, such as information directly from the 3D scene or the Azure CLU model. This will guide users to find the correct command instead of providing a misleading response to state that the system is incapable of completing the selection or editing task.

% \hl{[To be completed.]}

\section{Discussion}

This study highlights the promising potential of LLM-assisted interactive systems in guiding users towards more efficient multimodal interaction strategies, thereby improving user performance in typical interaction tasks such as scene editing in virtual reality.

\subsection{\added{LLM-Assisted Systems for VR}}

\added{Immersive environments present distinct spatial, multimodal, and latency requirements. For example, we found that users' reliance on both voice and visual feedback in the scene exposed limitations and areas of future work for such LLM-assisted systems to align natural language suggestions with visual cues in the 3D spatial context. This emphasizes the need to go beyond LLM applications for 2D interfaces to incorporate spatial reasoning capabilities in LLM-assisted systems for immersive content, and ensure alignment in different channels of natural language and visual information.}

\added{Additionally, our findings reveal how the immersive nature of VR amplifies certain characteristics in human-AI interaction. Hallucinated outputs from LLMs become more disruptive in immersive environments compared with traditional graphical user interfaces due to the higher cognitive load and conflicting visual information from the user's immediate spatial environment. These findings highlight the importance of addressing well-known issues of LLM applications, such as trust, multimodality, and uncertainty in the immersive domain.}

\subsection{\added{User Performance and Interaction Patterns}}

\paragraph{User Performance.}
Performance indicators, such as the number of remaining elemental editing steps and task completion time reported in \Cref{sec:overall-performance}, reveal how user performance significantly improved in Task 2 compared with Task 1. First,\deleted{ as shown by performance indicators,} the study exemplifies the impact of choosing the correct interaction strategy on task completion quality. While \textbf{431.3\%} more time was used on \textit{incremental exploration} as compared to \textit{bulk modification} in Task 1, 
users completed \textbf{66.38\%} more elemental editing steps with \textit{bulk modification} as compared with \textit{incremental exploration} in Task 1.
% \textit{bulk modification} resulted in a \textbf{66.38\%} reduction in the remaining elemental scene editing steps compared with the \textit{incremental exploration} strategy in Task 1. 
Second, the performance indicators reveal how LLM-assisted interactive systems \replaced{helped}{help} to guide users to select better interaction strategies which result in improved performance. In several cases for P12, P2, P6, and P5, the LLM-assisted scene editing system was able to give constructive feedback in response to user queries on the supported speech commands, available colors, available materials, and the most efficient way supported by the system for users to complete the scene editing task, with examples provided \replaced{in the Online Appendix}{below}. \deleted{This demonstrates how LLM-assisted interactive systems have the strong potential of handling various types of natural language user input and providing a response to the best of its customized knowledge base to guide users to improve their interaction strategy. } \added{However, the absence of a baseline (e.g., task completion without LLM assistance) limits our ability to definitively attribute these gains to the LLM, and results should be treated with caution.}

% \begin{quote}
%     \textit{``Can you make what I selected into another material?''} (P12, 2:18, Task 1)
    
%     \textit{``Yes, you can change the material of the selected objects. Simply use a speech command like 'Make it leather' or 'Make it cotton' to modify the material of the selected objects.''} (AssistVR)
% \end{quote}

% \begin{quote}
%     \textit{``What are the options for colours?''} (P2, 2:22, Task 1)
    
%     \textit{``You can change the color of an object to red; orange; yellow; green; blue; purple; brown; black; grey; white; red pattern; blue pattern; yellow pattern; purple pattern; or white pattern.''} (AssistVR)
% \end{quote}

% \begin{quote}
%     \textit{``What materials can I use?''} (P6, 6:29, Task 1)
    
%     \textit{``You can change the material of an object to plastic; cotton; leather; or concrete.''} (AssistVR)
% \end{quote}

% \begin{quote}
%     \textit{``What is the fastest way to change all the blue objects to purple?''} (P5, 4:58, Task 1)
    
%     \textit{``The fastest way to change all the blue objects to purple is to use speech commands. First, press Button A on the right controller and say 'Select all blue objects'. After selecting all blue objects, use the speech command 'Make them purple'.''} (AssistVR)
% \end{quote}

% The 

\paragraph{Interaction Patterns.}

Results from \Cref{sec:interaction-patterns} revealed certain interaction patterns\added{, offering insights on how individuals interact with LLMs in immersive environments. For example, users consistently prioritized visually salient features and objects with clear target states, suggesting that spatial and visual affordances play a significant role to guide AI-supported decision making. These findings highlight the need for LLM-assisted systems in immersive environments to align LLM suggestions with the user's focus and provide visual previews to align verbal commands with 3D outcomes.}

\added{The study also revealed that even when Task 1 and Task 2 explicitly encourage participants to find the most efficient way of scene editing, two out of twelve participants displayed rigid interaction behavior, adhering to a limited set of strategies regardless of system feedback. This observation reveals a potential barrier when users are encouraged to explore more optimal modes of interaction, and demonstrates the need for LLM-assisted systems to guide users towards exploiting system capabilities efficiently. 
}
\deleted{First, users tend to edit visually-distinguishable features, such as color properties, first before editing features with less distinct visual features, such as material properties. 
Second, when using the LLM-assisted interactive system for scene editing, participants preferred to edit objects with clear target states which could be edited through simple voice commands (i.e., `Make this purple', `Make them leather'), as opposed to objects with goal states that are difficult to enunciate and issue voice commands, such as the carpet.
Third, queries were mainly posed during incremental exploration as opposed to during bulk modification. More queries did not necessarily guide participants to find the bulk modification strategy. Instead, participants (i.e., P3--P6, P8) who tried some form of bulk modification in Task 1 and who posed some queries to the system were more successful in finding the most efficient strategy combination (bulk modification for most objects followed by incremental exploration to modify carpet pattern) in Task 2.
Finally, the study also revealed that participants tended to adhere to a single interaction strategy or a single pattern of interaction modalities. Even when Task 1 and Task 2 explicitly encourage participants to find the most efficient way of scene editing, P2 and P12 adhered to the incremental exploration strategy throughout the entire study and adhered to two interaction modality patterns: (1) modifying single objects directly through speech commands; or (2) using raycast to select one or several objects, then modifying the selected object(s) through a single speech command.}

% Stagnation in scene editing interaction modality and strategy

\subsection{Implications}

Results from this study shed light on design implications for future LLM-assisted interactive systems. While the study is conducted in a VR environment, design implications are applicable to 3D content applications in general, and even possibly applicable to general interactive systems where LLMs are involved. Based on results from the study, we formulate design implications as follows.

\textbf{First, effective use of multimodal input is critical for improving the user experience for LLM-assisted interactive systems.} \added{While multimodality has been widely-researched for immersive technology, our findings emphasize the need for dynamic integration of inputs such as controller, gesture, gaze, and verbal commands to support embodied interaction. For example, users frequently used verbal commands for high-level tasks but relied on controller input to finetune certain selections. This combination of high-level verbal commands combined with detailed-level controller or gesture input presents an opportunity for LLM-assisted systems to support more intuitive interactions in immersive environments and allows the system to disambiguate or refine user input which would otherwise be imprecise when only speech input is supported.} \deleted{In the study, many participants stumbled upon providing clear descriptions on how to edit the carpet. }
This corroborates findings from Liao et al.~\cite{liao2023ai} who state that interaction with LLM systems with only the natural language modality can be easily affected by subtle language \deleted{and communication }cues.
% This corroborates findings from \hl{XX et al.} who state that LLM-assisted interactive systems can face usability problems when the goal state is not or cannot be clearly specified using natural language. 
\deleted{To cope with this issue, Tsimpoukelli et al.~\cite{tsimpoukelli2021multimodal} allowed language models to support multimodal image and text input in addition to original natural language prompts. Wu et al. proposed NExT-GPT~\cite{wu2023next}, a general-purpose any-to-any MM-LLM system which supports inputs and outputs of any combination of text, image, video, and audio, demonstrating the importance of LLM-assisted interactive systems to leverage different information modalities to specify user intent. 
In the study, we also found cases where participants would like to use a combination of raycast selection and speech interaction to make queries about the selected object. For example, P7 asked, `Is this a leather object?' when using raycast to point at different objects.
Many participants also used raycast to point at the object of interest when asking questions, which further emphasizes the importance of leveraging multimodal information in interaction design. Associating with the comment by P7 on including eye gaze under `Multimodal Interaction' in \Cref{sec:post-exp-questionnaire}, this demonstrates the promise of using additional input such as eye gaze to contextualize a query.}

\textbf{Second, the design of LLM-assisted interactive systems should place special considerations on fostering user trust and improving user agency.} \added{Trust and agency are fundamental to LLM-assisted interactions. However, these concepts are expressed differently in VR due to its immersive nature. For example, users such as P2 expressed more trust in the incremental exploration strategy as it provided more visual confirmation. When P12 did not receive feedback on her command to select all blue objects, her sense of agency and control over the system through speech commands was affected. These observations inform future designs to allow users to preview the results of LLM-generated actions and incorporate reversible actions in addition to 2D visual displays such as the draggable panel implemented in \textsc{AssistVR} to reinforce a sense of control.} \deleted{Processing steps in LLM-assisted systems can easily become a black box to users as it is not evident how the system handles natural language input from the user. 
It was observed that when some participants first began using the system, they preferred to ask some simple questions to the system to verify that the system is processing their request as they expected, which is an important step for users to build trust with the system.}

\deleted{When asked about what made them to believe that the bulk modification strategy yielded correct results, some participants (P6, P8) confessed that they simply chose to trust the system but were not able to completely rule out the possibility that some editing steps might have gone wrong.}

\deleted{In the post-experience questionnaire, P2 and P12 used the incremental exploration strategy in both Task 1 and Task 2. P2 commented preferring this strategy over bulk modification because selecting individual objects provided more visual confirmation which consequently fostered trust with the system that it was making the correct editing step.
P12, on the other hand, chose the incremental exploration strategy because it provided more user agency and sense of control over the system. P12 commented that the system was not able to respond to her commands to select all blue objects, which affected her sense of agency and control over the system through speech commands.}

\deleted{While we used a draggable panel to assist users in verifying the correctness of the system output, we believe that there are many more possibilities to increase user trust and agency, for example, by overlaying or blending (future or already conducted) changes of 3D objects directly in the 3D scene and to allow for quick assertion and undo of previous actions.}

% 2. Reassuring trust: 
% P12: Hi can you tell me what I just selected?
% AssistVR: I'm sorry; but I can't determine what you just selected.

% P12: Can you make what I selected into another material?
% AssistVR: Yes, you can change the material of the selected objects. Simply use a speech command like 'Make it leather' or 'Make it cotton' to modify the material of the selected objects.

% Visual feedback and user agency

\textbf{{Finally}, LLM-assisted interactive systems should implement measures to convey the fundamental uncertainties that emerge from LLM interaction, such as hallucination.} \added{While the risk of hallucinations is well-recognized in AI research, our findings indicate that their impact is magnified in immersive environments due to their spatial and interactive nature. For example, hallucinated scene information can result in inconsistent responses. LLMs involved directly in scene editing (instead of attached to Unity post-processing scripts like the Azure CLU pipeline) can also result in erroneous spatial edits. These errors can all severely disrupt task flow and user immersion. To mitigate this, LLM-assisted systems for immersive content editing should explicitly convey uncertainty in LLM outputs through visual indicators (such as color-coded confidence levels) or audio cues. While LLMs inherently have limitations in hallucinating information~\cite{zhang2023siren}, it is important to signpost to the user when such information can be inaccurate or incomplete.} \deleted{In the study, there were some instances when hallucination occurred, which resulted in inaccurate or incomplete information provided by the system. In this regard, immersive environments could help to address the hallucination problem by providing additional contextual information to reduce uncertainty, which might help in minimizing the chances of hallucination compared to text-only input.}
% \begin{quote}
%     \textit{``Is there a bin in this room?''} (P9, 9:12, Task2)
    
%     \textit{``Based on the provided context, there is no information indicating the presence of a bin in the room.''} (AssistVR)
% \end{quote}

% \begin{quote}
%     \textit{``How many blue cotton items are there?''} (P9, 5:33)
    
%     \textit{``There is one blue cotton item, which is the pen holder.''} (AssistVR)
% \end{quote}
\deleted{P9 asked, ``Is there a bin in this room,'' and received the response ``Based on the provided context, there is no information indicating the presence of a bin in the room.'' 
In this case, there is a bin in the room but the system provided inaccurate information. P9 also asked, ``How many blue cotton items are there,'' and the system responded, ``There is one blue cotton item, which is the pen holder.'' In this example, there were several blue cotton items in the room, but the system provided incomplete information to list only the pen holder. While LLMs inherently have limitations in hallucinating information~\cite{zhang2023siren}, it is important to signpost to the user when such information can be inaccurate or incomplete.} 

% \textbf{\added{Finally, adaptive interaction support should also be considered to encourage exploration and account for individual differences.}} \added{The tendency of some users to rely on fixed interaction strategies highlights the importance of providing adaptive support mechanisms for LLM-assisted interactive systems. The system could offer real-time suggestions in visual or audio forms. Alternatively, features such as gamified elements or progress visualizations could incentivize users to experiment with different interaction modalities and strategies, which will help mitigate rigid interaction patterns for LLM-assisted systems. As the study also observed vastly different interaction patterns for different individuals, LLM-assisted systems should also include options for users to customize modalities based on their preferences and offer pathways for improvement by gradually exposing or leading them towards higher-level interaction techniques.}

% 3. Measures to cope with hallucination.

% The system can provide inaccurate or incomplete information.
% How many blue cotton items are there? (P9, 5:33)
% There is one blue cotton item; which is the pen holder.

% Is there a bin in this room? (P9, 9:12, Task2)
% Based on the provided context; there is no information indicating the presence of a bin in the room.

\section{Conclusion and Future Outlook}

This work has provided an analysis of user interaction patterns and strategies with LLM-assisted interactive systems through an example scene editing task in virtual reality. As evidenced by the results in \Cref{sec:overall-performance}, LLM-assisted interactive systems have the potential to guide users to find more effective and efficient interaction strategies and improve task performance with very limited external guidance. Results from the post-experience questionnaire corroborate findings in prior work~\cite{kurai2024magicitem, giunchi2024dreamcodevr} on the strengths of LLM-assisted interactive systems for immersive content in perceived workload, usability, and user experience. Based on post-experience questionnaire comments, we summarize design considerations for LLM-assisted interactive systems in terms of multimodal interaction, user trust, user agency, and appropriate feedback to cope with uncertainty and hallucination. We also summarize interaction patterns such as the fact that visually distinguishable features tend to be edited first, and objects with an obscure goal state tend to be edited last. Interaction patterns further reveal how participants were able to improve their strategy through interaction with the system. Based on these qualitative and quantitative observations, we proposed a set of design implications for LLM-assisted interactive systems.

Novelty effects possibly inflated usability perceptions and thus results have to be treated with caution and we encourage replication efforts. \added{We also acknowledge limitations in our task design. For example, our tasks do not fully capture or analyze ambiguous user input cases typical of speech interfaces. Our tasks also do not take into account complex high-level editing requirements from the user. The current tasks on color and material modification and the \textsc{AssistVR} system presented in the paper lack the versatility to cater to various design requirements, which will be addressed in future work. Nevertheless,} \replaced{this}{This} study provides a promising outlook for LLM-assisted interactive systems and provides a reference for future work on interaction analysis of LLM-assisted systems. We envision that these interaction pattern findings and design implications will be applicable to LLM-assisted interactive systems in general to guide a broad range of future designs in VR and beyond.

% put future work (frame limitations positively) here

% \added{
\section*{Supplemental Materials}
\label{sec:supplemental_materials}
% }

\added{
Supplementary materials can be found in the Online Appendix at \url{https://osf.io/2hmnd/}. This includes: (1) The utterance training data for Azure CLU, the prompts for GPT-4o, (2) Verbal instructions given to participants before the practice trial, and before both tasks and both task types, and (3) Selected conversation histories between participants and \textsc{AssistVR} grouped into themes.
}

%\section*{Supplemental Materials}
%\label{sec:supplemental_materials}

%\section*{Figure Credits}
%\label{sec:figure_credits}

\acknowledgments{
\replaced{Junlong Chen is supported by the China Scholarship Council and Cambridge Trust.}{The authors wish to thank A, B, and C. This work was supported in part by a grant from XYZ.}}

\bibliographystyle{abbrv-doi}

\bibliography{template}

@inproceedings{li2024discene,
author = {Li, Xiao-Lei and Li, Haodong and Chen, Hao-Xiang and Mu, Tai-Jiang and Hu, Shi-Min},
title = {{DIScene: Object Decoupling and Interaction Modeling for Complex Scene Generation}},
year = {2024},
isbn = {9798400711312},
publisher = {Association for Computing Machinery},
address = {New York, NY, USA},
url = {https://doi.org/10.1145/3680528.3687589},
doi = {10.1145/3680528.3687589},
booktitle = {SIGGRAPH Asia 2024 Conference Papers},
articleno = {101},
numpages = {12},
location = {
},
series = {SA '24}
}

@inproceedings{de2024llmr,
  title={{LLMR: Real-time prompting of interactive worlds using large language models}},
  author={De La Torre, Fernanda and Fang, Cathy Mengying and Huang, Han and Banburski-Fahey, Andrzej and Amores Fernandez, Judith and Lanier, Jaron},
  booktitle={Proceedings of the CHI Conference on Human Factors in Computing Systems},
  pages={1--22},
  year={2024}
}

@article{OpenAI_2024, 
note = {Available at \url{https://openai.com/index/hello-gpt-4o}},
journal={Hello GPT-4o}, 
author={OpenAI}, 
year={2024}, 
month={May}}

@article{achiam2023gpt,
  title={{GPT-4 Technical Report}},
  author={Achiam, Josh and Adler, Steven and Agarwal, Sandhini and Ahmad, Lama and Akkaya, Ilge and Aleman, Florencia Leoni and Almeida, Diogo and Altenschmidt, Janko and Altman, Sam and Anadkat, Shyamal and others},
  journal={arXiv preprint arXiv:2303.08774},
  year={2023}
}

@incollection{hart1988development,
title = {{Development of NASA-TLX (Task Load Index): Results of Empirical and Theoretical Research}},
editor = {Peter A. Hancock and Najmedin Meshkati},
series = {Advances in Psychology},
publisher = {North-Holland},
volume = {52},
pages = {139-183},
year = {1988},
booktitle = {Human Mental Workload},
issn = {0166-4115},
DOI = {https://doi.org/10.1016/S0166-4115(08)62386-9},
url = {https://www.sciencedirect.com/science/article/pii/S0166411508623869},
author = {Sandra G. Hart and Lowell E. Staveland}
}

@article{brooke1996sus,
  title={{SUS-A quick and dirty usability scale}},
  author={Brooke, John and others},
  journal={Usability evaluation in industry},
  volume={189},
  number={194},
  pages={4--7},
  year={1996},
  publisher={London, England}
}

@article{zhang2023siren,
  title={{Siren's Song in the AI Ocean: A Survey on Hallucination in Large Language Models}},
  author={Zhang, Yue and Li, Yafu and Cui, Leyang and Cai, Deng and Liu, Lemao and Fu, Tingchen and Huang, Xinting and Zhao, Enbo and Zhang, Yu and Chen, Yulong and others},
  journal={arXiv preprint arXiv:2309.01219},
  year={2023}
}

@article{wu2023next,
  title={{NExT-GPT: Any-to-Any Multimodal LLM}},
  author={Wu, Shengqiong and Fei, Hao and Qu, Leigang and Ji, Wei and Chua, Tat-Seng},
  journal={arXiv preprint arXiv:2309.05519},
  year={2023}
}

@article{liao2023ai,
  title={{AI Transparency in the Age of LLMs: A Human-Centered Research Roadmap}},
  author={Liao, Q Vera and Vaughan, Jennifer Wortman},
  journal={arXiv preprint arXiv:2306.01941},
  pages={5368--5393},
  year={2023},
  publisher={no}
}

@inproceedings{gpt3,
 author = {Brown, Tom and Mann, Benjamin and Ryder, Nick and Subbiah, Melanie and Kaplan, Jared D and Dhariwal, Prafulla and Neelakantan, Arvind and Shyam, Pranav and Sastry, Girish and Askell, Amanda and Agarwal, Sandhini and Herbert-Voss, Ariel and Krueger, Gretchen and Henighan, Tom and Child, Rewon and Ramesh, Aditya and Ziegler, Daniel and Wu, Jeffrey and Winter, Clemens and Hesse, Chris and Chen, Mark and Sigler, Eric and Litwin, Mateusz and Gray, Scott and Chess, Benjamin and Clark, Jack and Berner, Christopher and McCandlish, Sam and Radford, Alec and Sutskever, Ilya and Amodei, Dario},
 booktitle = {Advances in Neural Information Processing Systems},
 editor = {H. Larochelle and M. Ranzato and R. Hadsell and M.F. Balcan and H. Lin},
 pages = {1877--1901},
 publisher = {Curran Associates, Inc.},
 title = {Language Models are Few-Shot Learners},
 url = {https://proceedings.neurips.cc/paper_files/paper/2020/file/1457c0d6bfcb4967418bfb8ac142f64a-Paper.pdf},
 volume = {33},
 year = {2020}
}

@article{konenkov2024vrgpt,
  title={{VR-GPT: Visual Language Model for Intelligent Virtual Reality Applications}},
  author={Konenkov, Mikhail and Lykov, Artem and Trinitatova, Daria and Tsetserukou, Dzmitry},
  journal={arXiv preprint arXiv:2405.11537},
  year={2024}
}

@inproceedings{giunchi2024dreamcodevr,
  title={{DreamCodeVR: Towards Democratizing Behavior Design in Virtual Reality with Speech-Driven Programming}},
  author={Giunchi, Daniele and Numan, Nels and Gatti, Elia and Steed, Anthony},
  booktitle={2024 IEEE Conference Virtual Reality and 3D User Interfaces (VR)},
  pages={579--589},
  year={2024},
  organization={IEEE}
}

@article{tsimpoukelli2021multimodal,
  title={{Multimodal Few-Shot Learning with Frozen Language Models}},
  author={Tsimpoukelli, Maria and Menick, Jacob L and Cabi, Serkan and Eslami, SM and Vinyals, Oriol and Hill, Felix},
  journal={Advances in Neural Information Processing Systems},
  volume={34},
  pages={200--212},
  year={2021}
}

@article{schrepp2017design,
  title={{Design and Evaluation of a Short Version of the User Experience Questionnaire (UEQ-S)}},
  author={Schrepp, Martin and Hinderks, Andreas and Thomaschewski, J{\"o}rg},
  journal={International Journal of Interactive Multimedia and Artificial Intelligence, 4 (6), 103-108.},
  year={2017},
  publisher={UNIR}
}

@inproceedings{aghel2024people,
  title={{How People Prompt Generative AI to Create Interactive VR Scenes}},
  author={Aghel Manesh, Setareh and Zhang, Tianyi and Onishi, Yuki and Hara, Kotaro and Bateman, Scott and Li, Jiannan and Tang, Anthony},
  booktitle={Proceedings of the 2024 ACM Designing Interactive Systems Conference},
  pages={2319--2340},
  year={2024}
}

@article{rakkolainen2021technologies,
  title={{Technologies for Multimodal Interaction in Extended Reality—A Scoping Review}},
  author={Rakkolainen, Ismo and Farooq, Ahmed and Kangas, Jari and Hakulinen, Jaakko and Rantala, Jussi and Turunen, Markku and Raisamo, Roope},
  journal={Multimodal Technologies and Interaction},
  volume={5},
  number={12},
  pages={81},
  year={2021},
  publisher={MDPI}
}

@article{spittle2022review,
  title={{A Review of Interaction Techniques for Immersive Environments}},
  author={Spittle, Becky and Frutos-Pascual, Maite and Creed, Chris and Williams, Ian},
  journal={IEEE Transactions on Visualization and Computer Graphics},
  volume={29},
  number={9},
  pages={3900--3921},
  year={2022},
  publisher={IEEE}
}

@article{williams2020understanding,
  title={{Understanding Gesture and Speech Multimodal Interactions for Manipulation Tasks in Augmented Reality Using Unconstrained Elicitation}},
  author={Williams, Adam S and Ortega, Francisco R},
  journal={Proceedings of the ACM on Human-Computer Interaction},
  volume={4},
  number={ISS},
  pages={1--21},
  year={2020},
  publisher={ACM New York, NY, USA}
}

@inproceedings{zhou2022eliciting,
  title={{Eliciting Multimodal Gesture+Speech Interactions in a Multi-Object Augmented Reality Environment}},
  author={Zhou, Xiaoyan and Williams, Adam Sinclair and Ortega, Francisco Raul},
  booktitle={Proceedings of the 28th ACM Symposium on Virtual Reality Software and Technology},
  pages={1--10},
  year={2022}
}

@article{plopski2022eye,
  title={{The Eye in Extended Reality: A Survey on Gaze Interaction and Eye Tracking in Head-worn Extended Reality}},
  author={Plopski, Alexander and Hirzle, Teresa and Norouzi, Nahal and Qian, Long and Bruder, Gerd and Langlotz, Tobias},
  journal={ACM Computing Surveys (CSUR)},
  volume={55},
  number={3},
  pages={1--39},
  year={2022},
  publisher={ACM New York, NY}
}

@inproceedings{lee2024gazepointar,
  title={{GazePointAR: A Context-Aware Multimodal Voice Assistant for Pronoun Disambiguation in Wearable Augmented Reality}},
  author={Lee, Jaewook and Wang, Jun and Brown, Elizabeth and Chu, Liam and Rodriguez, Sebastian S and Froehlich, Jon E},
  booktitle={Proceedings of the CHI Conference on Human Factors in Computing Systems},
  pages={1--20},
  year={2024}
}

@inproceedings{zimmerer2020finally,
  title={{Finally on Par?! Multimodal and Unimodal Interaction for Open Creative Design Tasks in Virtual Reality}},
  author={Zimmerer, Chris and Wolf, Erik and Wolf, Sara and Fischbach, Martin and Lugrin, Jean-Luc and Latoschik, Marc Erich},
  booktitle={Proceedings of the 2020 International Conference on Multimodal Interaction},
  pages={222--231},
  year={2020}
}

@inproceedings{rodriguez2024artists,
  title={{An Artists' Perspectives on Natural Interactions for Virtual Reality 3D Sketching}},
  author={Rodriguez, Richard and Sullivan, Brian T and Barrera Machuca, Mayra Donaji and Batmaz, Anil Ufuk and Tornatzky, Cyane and Ortega, Francisco R},
  booktitle={Proceedings of the CHI Conference on Human Factors in Computing Systems},
  pages={1--20},
  year={2024}
}

@inproceedings{jiang2023beyond,
  title={{Beyond Audio Description: Exploring 360° Video Accessibility with Blind and Low Vision Users Through Collaborative Creation}},
  author={Jiang, Lucy and Phutane, Mahika and Azenkot, Shiri},
  booktitle={Proceedings of the 25th international ACM SIGACCESS conference on computers and accessibility},
  pages={1--17},
  year={2023}
}

@inproceedings{bozkir2024embedding,
  title={{Embedding Large Language Models into Extended Reality: Opportunities and Challenges for Inclusion, Engagement, and Privacy}},
  author={Bozkir, Efe and {\"O}zdel, S{\"u}leyman and Lau, Ka Hei Carrie and Wang, Mengdi and Gao, Hong and Kasneci, Enkelejda},
  booktitle={Proceedings of the 6th ACM Conference on Conversational User Interfaces},
  pages={1--7},
  year={2024}
}

@incollection{he2024enhancing,
  title={{Enhancing Narratives with SayMotion's text-to-3D animation and LLMs}},
  author={He, Kevin and Lapham, Annette and Li, Zenan},
  booktitle={ACM SIGGRAPH 2024 Real-Time Live!},
  pages={1--2},
  year={2024}
}

@inproceedings{manfredi2023mixed,
  title={{A Mixed Reality Approach for Innovative Pair Programming Education with a Conversational AI Virtual Avatar}},
  author={Manfredi, Gilda and Erra, Ugo and Gilio, Gabriele},
  booktitle={Proceedings of the 27th International Conference on Evaluation and Assessment in Software Engineering},
  pages={450--454},
  year={2023}
}

@inproceedings{song2023expanded,
  title={{From Expanded Cinema to Extended Reality: How AI Can Expand and Extend Cinematic Experiences}},
  author={Song, Junrong and Wang, Bingyuan and Wang, Zeyu and Yip, David Kei-Man},
  booktitle={Proceedings of the 16th International Symposium on Visual Information Communication and Interaction},
  pages={1--5},
  year={2023}
}

@article{ma2024llms,
  title={{When LLMs step into the 3D World: A Survey and Meta-Analysis of 3D Tasks via Multi-modal Large Language Models}},
  author={Ma, Xianzheng and Bhalgat, Yash and Smart, Brandon and Chen, Shuai and Li, Xinghui and Ding, Jian and Gu, Jindong and Chen, Dave Zhenyu and Peng, Songyou and Bian, Jia-Wang and others},
  journal={arXiv preprint arXiv:2405.10255},
  year={2024}
}

@InProceedings{huang2022language,
  title = 	 {{Language Models as Zero-Shot Planners: Extracting Actionable Knowledge for Embodied Agents}},
  author =       {Huang, Wenlong and Abbeel, Pieter and Pathak, Deepak and Mordatch, Igor},
  booktitle = 	 {Proceedings of the 39th International Conference on Machine Learning},
  pages = 	 {9118--9147},
  year = 	 {2022},
  editor = 	 {Chaudhuri, Kamalika and Jegelka, Stefanie and Song, Le and Szepesvari, Csaba and Niu, Gang and Sabato, Sivan},
  volume = 	 {162},
  series = 	 {Proceedings of Machine Learning Research},
  month = 	 {17--23 Jul},
  publisher =    {PMLR},
  url = 	 {https://proceedings.mlr.press/v162/huang22a.html}
}

@article{huang2022inner,
  title={{Inner Monologue: Embodied Reasoning through Planning with Language Models}},
  author={Huang, Wenlong and Xia, Fei and Xiao, Ted and Chan, Harris and Liang, Jacky and Florence, Pete and Zeng, Andy and Tompson, Jonathan and Mordatch, Igor and Chebotar, Yevgen and others},
  journal={arXiv preprint arXiv:2207.05608},
  year={2022}
}

@article{ahn2022can,
  title={{Do As I Can, Not As I Say: Grounding Language in Robotic Affordances}},
  author={Ahn, Michael and Brohan, Anthony and Brown, Noah and Chebotar, Yevgen and Cortes, Omar and David, Byron and Finn, Chelsea and Fu, Chuyuan and Gopalakrishnan, Keerthana and Hausman, Karol and others},
  journal={arXiv preprint arXiv:2204.01691},
  year={2022}
}

@article{roberts2022steps,
  title={{Steps towards prompt-based creation of virtual worlds}},
  author={Roberts, Jasmine and Banburski-Fahey, Andrzej and Lanier, Jaron},
  journal={arXiv preprint arXiv:2211.05875},
  year={2022}
}

@article{rabsahl2023symbolic,
  title={{Symbolic Event Visualization for Analyzing User Input and Behavior of Augmented Reality Sessions}},
  author={Rabsahl, Solveig and Satzger, Thomas and Kalamkar, Snehanjali and Grubert, Jens and Beck, Fabian},
  year={2023},
  publisher={Otto-Friedrich-Universit{\"a}t}
}

@inproceedings{scholz2024classifying,
  title={{Classifying User Roles in Online News Forums: A Model for User Interaction and Behavior Analysis}},
  author={Scholz, Felix and Kolb, Thomas Elmar and Neidhardt, Julia},
  booktitle={Adjunct Proceedings of the 32nd ACM Conference on User Modeling, Adaptation and Personalization},
  pages={240--249},
  year={2024}
}

@article{wright2000analyzing,
  title={{Analyzing Human-Computer Interaction as Distributed Cognition: The Resources Model}},
  author={Wright, Peter C and Fields, Robert E and Harrison, Michael D},
  journal={Human-Computer Interaction},
  volume={15},
  number={1},
  pages={1--41},
  year={2000},
  publisher={Taylor \& Francis}
}

@inproceedings{guo2024investigating,
  title={{Investigating Interaction Modes and User Agency in Human-LLM Collaboration for Domain-Specific Data Analysis}},
  author={Guo, Jiajing and Mohanty, Vikram and Piazentin Ono, Jorge H and Hao, Hongtao and Gou, Liang and Ren, Liu},
  booktitle={Extended Abstracts of the CHI Conference on Human Factors in Computing Systems},
  pages={1--9},
  year={2024}
}

@article{jebeli2024quantifying,
  title={{Quantifying the Quality of Parent-Child Interaction Through Machine-Learning Based Audio and Video Analysis: Towards a Vision of AI-assisted Coaching Support for Social Workers}},
  author={Jebeli, Atefeh and Chen, Lujie Karen and Guerrerio, Katherine and Papparotto, Sophia and Berlin, Lisa and Harden, Brenda Jones},
  journal={ACM Journal on Computing and Sustainable Societies},
  volume={2},
  number={1},
  pages={1--21},
  year={2024},
  publisher={ACM New York, NY}
}

@inproceedings{trippas2024users,
  title={{What do Users Really Ask Large Language Models? An Initial Log Analysis of Google Bard Interactions in the Wild}},
  author={Trippas, Johanne R and Al Lawati, Sara Fahad Dawood and Mackenzie, Joel and Gallagher, Luke},
  booktitle={Proceedings of the 47th International ACM SIGIR Conference on Research and Development in Information Retrieval},
  pages={2703--2707},
  year={2024}
}

@article{beyan2023co,
  title={{Co-Located Human--Human Interaction Analysis Using Nonverbal Cues: A Survey}},
  author={Beyan, Cigdem and Vinciarelli, Alessandro and Bue, Alessio Del},
  journal={ACM Computing Surveys},
  volume={56},
  number={5},
  pages={1--41},
  year={2023},
  publisher={ACM New York, NY}
}

@inproceedings{feit2016we,
  title={{How We Type: Movement Strategies and Performance in Everyday Typing}},
  author={Feit, Anna Maria and Weir, Daryl and Oulasvirta, Antti},
  booktitle={Proceedings of the 2016 CHI Conference on Human Factors in Computing Systems},
  pages={4262--4273},
  numpages = {12},
  year={2016}
}

@inproceedings{foy2021understanding,
  title={{Understanding, Detecting and Mitigating the Effects of Coactivations in Ten-Finger Mid-Air Typing in Virtual Reality}},
  author={Foy, Conor R and Dudley, John J and Gupta, Aakar and Benko, Hrvoje and Kristensson, Per Ola},
  booktitle={Proceedings of the 2021 CHI conference on Human Factors in Computing Systems},
  pages={1--11},
  year={2021}
}

@inproceedings{dudley2019performance,
  title={{Performance Envelopes of Virtual Keyboard Text Input Strategies in Virtual Reality}},
  author={Dudley, John and Benko, Hrvoje and Wigdor, Daniel and Kristensson, Per Ola},
  booktitle={2019 IEEE International Symposium on Mixed and Augmented Reality (ISMAR)},
  pages={289--300},
  year={2019},
  organization={IEEE}
}

@article{kurai2024magicitem,
  title={{MagicItem: Dynamic Behavior Design of Virtual Objects with Large Language Models in a Consumer Metaverse Platform}},
  author={Kurai, Ryutaro and Hiraki, Takefumi and Hiroi, Yuichi and Hirao, Yutaro and Perusquia-Hernandez, Monica and Uchiyama, Hideaki and Kiyokawa, Kiyoshi},
  journal={arXiv preprint arXiv:2406.13242},
  year={2024}
}

@inproceedings{bolt1980put,
  title={{“Put-that-there” Voice and gesture at the graphics interface}},
  author={Bolt, Richard A},
  booktitle={Proceedings of the 7th annual conference on Computer graphics and interactive techniques},
  pages={262--270},
  year={1980}
}

@inproceedings{kaiser2003mutual,
  title={{Mutual Disambiguation of 3D Multimodal Interaction in Augmented and Virtual Reality}},
  author={Kaiser, Ed and Olwal, Alex and McGee, David and Benko, Hrvoje and Corradini, Andrea and Li, Xiaoguang and Cohen, Phil and Feiner, Steven},
  booktitle={Proceedings of the 5th International Conference on Multimodal Interfaces},
  pages={12--19},
  year={2003}
}

@inproceedings{bergstrom2021evaluate,
  title={{How to Evaluate Object Selection and Manipulation in VR? Guidelines from 20 Years of Studies}},
  author={Bergstr{\"o}m, Joanna and Dalsgaard, Tor-Salve and Alexander, Jason and Hornb{\ae}k, Kasper},
  booktitle={proceedings of the 2021 CHI conference on human computing systems},
  pages={1--20},
  year={2021}
}

@article{yu2024object,
  title={{Object Selection and Manipulation in VR Headsets: Research Challenges, Solutions, and Success Measurements}},
  author={Yu, Difeng and Dingler, Tilman and Velloso, Eduardo and Goncalves, Jorge},
  journal={ACM Computing Surveys},
  year={2024},
  publisher={ACM New York, NY}
}

@article{argelaguet2013survey,
  title={{A survey of 3D object selection techniques for virtual environments}},
  author={Argelaguet, Ferran and Andujar, Carlos},
  journal={Computers \& Graphics},
  volume={37},
  number={3},
  pages={121--136},
  year={2013},
  publisher={Elsevier}
}

@article{guest2012introduction,
  title={{Introduction to Applied Thematic Analysis}},
  author={Guest, Gregory and MacQueen, Kathleen M and Namey, Emily E},
  journal={Applied Thematic Analysis},
  volume={3},
  number={20},
  pages={1--21},
  year={2012}
}
\balance
\end{document}